\documentclass[superscriptaddress,aps,pra,nofootinbib,twocolumn,notitlepage,10pt]{revtex4-1}
\usepackage{graphicx}
\usepackage{amsmath}
\usepackage{amssymb}
\usepackage{amsthm}
\usepackage{comment}
\usepackage{placeins}
\usepackage[caption=false]{subfig}
\usepackage[colorlinks,breaklinks=true]{hyperref}
\usepackage[all]{hypcap}
\usepackage{tikz}
\usepackage{verbatim}
\usepackage{subfig}
\usetikzlibrary{arrows}
\usepackage{units}
\usepackage{soul}

\usepackage{algorithmicx,algorithm,algpseudocode}

\newcommand{\eq}[1]{Eq.~\hyperref[eq:#1]{(\ref*{eq:#1})}}
\renewcommand{\sec}[1]{\hyperref[sec:#1]{Section~\ref*{sec:#1}}}
\DeclareRobustCommand{\app}[1]{\hyperref[app:#1]{Appendix~\ref*{app:#1}}}
\newcommand{\tab}[1]{\hyperref[tab:#1]{Table~\ref*{tab:#1}}}
\newcommand{\fig}[1]{\hyperref[fig:#1]{Figure~\ref*{fig:#1}}}
\newcommand{\figa}[2]{\hyperref[fig:#1]{Figure~\ref*{fig:#1}#2}}
\newcommand{\figx}[2]{\hyperref[fig:#1]{Figure~\ref*{fig:#1}(#2)}}
\newcommand{\thm}[1]{\hyperref[thm:#1]{Theorem~\ref*{thm:#1}}}
\newcommand{\lem}[1]{\hyperref[lem:#1]{Lemma~\ref*{lem:#1}}}
\newcommand{\cor}[1]{\hyperref[cor:#1]{Corollary~\ref*{cor:#1}}}
\newcommand{\defn}[1]{\hyperref[def:#1]{Definition~\ref*{def:#1}}}
\newcommand{\alg}[1]{\hyperref[alg:#1]{Algorithm~\ref*{alg:#1}}}
\newcommand{\clm}[1]{\hyperref[claim:#1]{Claim~\ref*{claim:#1}}}
\newcommand{\step}[1]{\hyperref[step:#1]{Step~\ref*{step:#1}}}

\def\bra#1{\mathinner{\langle{#1}|}}
\def\ket#1{\mathinner{|{#1}\rangle}}
\newcommand{\braket}[2]{\langle #1|#2\rangle}
\newcommand{\proj}[1]{\ket{#1}\!\!\bra{#1}}

\input{Qcircuit}

\begin{document}

\title{Improved Techniques for Preparing Eigenstates of Fermionic Hamiltonians}

\date{\today}
\author{Dominic W. Berry}
\affiliation{Department of Physics and Astronomy, Macquarie University, Sydney, NSW 2109, Australia}
\author{M\'{a}ria Kieferov\'{a}}
\affiliation{Department of Physics and Astronomy, Macquarie University, Sydney, NSW 2109, Australia}
\affiliation{Institute for Quantum Computing and Department of Physics and Astronomy, University of Waterloo, Waterloo, ON N2L 3G1, Canada}
\author{Artur Scherer}
\affiliation{Department of Physics and Astronomy, Macquarie 
University, Sydney, NSW 2109, Australia}
\author{Yuval R. Sanders}
\affiliation{Department of Physics and Astronomy, Macquarie University, Sydney, NSW 2109, Australia}
\author{Guang Hao Low}
\affiliation{Microsoft Research, Redmond, WA 98052, United States of America}
\author{Nathan Wiebe}
\affiliation{Microsoft Research, Redmond, WA 98052, United States of America}
\author{Craig Gidney}
\affiliation{Google Inc., Santa Barbara, CA 93117, United States of America}
\author{Ryan Babbush}
\email[Corresponding author: ]{babbush@google.com}
\affiliation{Google Inc., Venice, CA 90291, United States of America}

\begin{abstract}
Modeling low energy eigenstates of fermionic systems can provide insight into chemical reactions and material properties and is one of the most anticipated applications of quantum computing. We present three techniques for reducing the cost of preparing fermionic Hamiltonian eigenstates using phase estimation. First, we report a polylogarithmic-depth quantum algorithm for antisymmetrizing the initial states required for simulation of fermions in first quantization. This is an exponential improvement over the previous state-of-the-art. Next, we show how to reduce the overhead due to repeated state preparation in phase estimation when the goal is to prepare the ground state to high precision and one has knowledge of an upper bound on the ground state energy that is less than the excited state energy (often the case in quantum chemistry). Finally, we explain how one can perform the time evolution necessary for the phase estimation based preparation of Hamiltonian eigenstates with exactly zero error by using the recently introduced qubitization procedure.
\end{abstract}
\maketitle

\section{Introduction}

One of the most important applications of quantum simulation (and of quantum computing in general) is the Hamiltonian simulation based solution of the electronic structure problem. The ability to accurately model ground states of fermionic systems would have significant implications for many areas of chemistry and materials science and could enable the in silico design of new solar cells, batteries, catalysts, pharmaceuticals, etc.\ \cite{Mueck2015,Mohseni2017}. The most rigorous approaches to solving this problem involve using the quantum phase estimation algorithm \cite{Kitaev1995} to project to molecular ground states starting from a classically guessed state \cite{Aspuru-Guzik2005}. Beyond applications in chemistry, one might want to prepare fermionic eigenstates in order to simulate quantum materials \cite{Bauer2015b} including models of high-temperature superconductivity \cite{Jiang2017}.

In the procedure introduced by Abrams and Lloyd \cite{Abrams1999}, one first initializes the system in some efficient-to-prepare initial state $\ket{\varphi}$ which has appreciable support on the desired eigenstate $\ket{k}$ of Hamiltonian $H$. One then uses quantum simulation to construct a unitary operator that approximates time evolution under $H$. With these ingredients, standard phase estimation techniques invoke controlled application of powers of $U(\tau)=e^{-iH\tau}$. With probability $\alpha_k = |\braket{\varphi}{k}|^2$, the output is then an estimate of the corresponding eigenvalue ${E}_{k}$ with standard deviation $\sigma_{E_{k}} = O((\tau M)^{-1})$, where $M$ is the total number of applications of $U(\tau)$. The synthesis of $e^{-iH\tau}$ is typically performed using digital quantum simulation algorithms, such as by Lie-Trotter product formulas~\cite{Berry2007}, truncated Taylor series~\cite{Berry2015}, or quantum signal processing~\cite{Low2017}.

Since the proposal by Abrams and Lloyd \cite{Abrams1999}, algorithms for time-evolving fermionic systems have improved substantially \cite{Ortiz2001,Whitfield2010,Hastings2015,Poulin2014,Sugisaki2016,Motzoi2017,BabbushLow,Kivlichan2017}. Innovations that are particularly relevant to this paper  include the use of first quantization to reduce spatial overhead \cite{Boghosian1998,Boghosian1998b,Zalka1998,Kassal2008,Toloui2013,Kivlichan2016} from $O(N)$ to $O(\eta \log N)$ where $\eta$ is number of particles and $N \gg \eta$ is number of single-particle basis functions (e.g.\ molecular orbitals or plane waves), and the
use of post-Trotter methods to reduce the scaling with time-evolution error from $O(\textrm{poly}(1/\epsilon))$ to $O(\textrm{polylog}(1/\epsilon))$ \cite{BabbushSparse1,BabbushSparse2,Kivlichan2016}.  The algorithm of \cite{Kivlichan2016} makes use of both of these techniques to enable the most efficient first quantized quantum simulation of electronic structure in the literature.

Unlike second quantized simulations which necessarily scale polynomially in $N$, first quantized simulation offers the possibility of achieving total gate complexity $O(\textrm{poly}(\eta)\, \textrm{polylog}(N, 1/\epsilon))$. This is important because the convergence of basis set discretization error is limited by resolution of the electron-electron cusp \cite{Kato1957}, which cannot be resolved faster than $O(1/N)$ using any single-particle basis expansion. Thus, whereas the cost of refining second quantized simulations to within $\delta$ of the continuum basis limit is necessarily $O(\textrm{poly}(1/\delta))$, first quantization offers the possibility of suppressing basis set errors as $O(\textrm{polylog}(1/\delta))$, providing essentially arbitrarily precise representations.

In second quantized simulations of fermions the wavefunction encodes an antisymmetric fermionic system, but the qubit representation of that wavefunction is not necessarily antisymmetric. Thus, in second quantization it is necessary that operators act on the encoded wavefunction in a way that enforces the proper exchange statistics. This is the purpose of second quantized fermion mappings such as those explored in \cite{Somma2002,Seeley2012,Tranter2015,Bravyi2017,Havlicek2017,Setia2017,Steudtner2017}. By contrast, the distinguishing feature of first quantized simulations is that the antisymmetry of the encoded system must be enforced directly in the qubit representation of the wavefunction. This often simplifies the task of Hamiltonian simulation but complicates the initial state preparation.

In first quantization there are typically $\eta$ different registers of size $\log N$ (where $\eta$ is the number of particles and $N$ is number of spin-orbitals) encoding integers indicating the indices of occupied orbitals. As only $\eta$ of the $N$ orbitals are occupied, with $\eta \log N$ qubits one can specify an arbitrary configuration. To perform simulations in first quantization, one typically requires that the initial state $\ket{\varphi}$ is antisymmetric under the exchange of any two of the $\eta$ registers. Prior work presented a procedure for preparing such antisymmetric states  with gate complexity scaling as $\widetilde{O}(\eta^2)$ \cite{Abrams1994,Abrams1997}.

In \sec{symm} we provide a general approach for antisymmetrizing states via sorting networks. The circuit size is $O(\eta \log^c \eta \log N)$ and the depth is $O(\log^c \eta \log \log N)$, where the value of $c\geq 1$ depends on the choice of sorting network (it can be $1$, albeit with a large multiplying factor).
In terms of the circuit depth, these results improve exponentially over prior implementations ~\cite{Abrams1997,Abrams1994}.
They also improve polynomially on the total number of gates needed.
We also discuss an alternative approach, a quantum variant of the Fisher-Yates shuffle, which avoids sorting, and achieves a size-complexity of $O (\eta^2 \log N)$ with lower spatial overhead than the sort-based methods.

Once the initial state $\ket{\varphi}$ has been prepared, it typically will not be exactly the ground state desired. In the usual approach, one would perform phase estimation repeatedly until the ground state is obtained, giving an overhead scaling inversely with the initial state overlap. In \sec{rejection} we propose a strategy for reducing this cost, by initially performing the estimation with only enough precision to eliminate excited states.

In \sec{no_error} we explain how qubitization \cite{Low2016} provides a unitary sufficient for phase estimation purposes with exactly zero error (provided a gate set consisting of an entangling gate and arbitrary single-qubit rotations). This improves over proposals to perform the time evolution unitary with post-Trotter methods at cost scaling as $O(\textrm{polylog}(1/\epsilon))$. We expect that a combination of these strategies will enable quantum simulations of fermions similar to the proposal of \cite{Kivlichan2016} with substantially fewer T gates than any method suggested in prior literature.

\section{Exponentially Faster Antisymmetrization}
\label{sec:symm}

Here we present our algorithm for imposing fermionic exchange symmetry on a sorted, repetition-free quantum array $\texttt{target}$.
Specifically, the result of this procedure is to perform the transformation
\begin{align}
\ket{r_1 \cdots r_\eta} \mapsto
\sum_{\sigma \in S_\eta} \left(-1\right)^{\pi\left(\sigma\right)}\ket{\sigma\left(r_1, \cdots, r_\eta\right)}
\end{align}
where $\pi(\sigma)$ is the parity of the permutation $\sigma$, and we require for the initial state that $r_p < r_{p+1}$ (necessary for this procedure to be unitary).
Although we describe the procedure for a single input $\ket{r_1 \cdots r_\eta}$, our algorithm may be applied to any superposition of such states.

Our approach is a modification of that proposed in Refs.~\cite{Abrams1997,Abrams1994};
namely, to apply the reverse of a sort to a sorted quantum array.
Whereas Refs.~\cite{Abrams1997,Abrams1994} provide a gate count of $O (\eta^2 (\log N)^2)$,
we can report a gate count of $O (\eta \log \eta \log N)$ and a runtime of $O (\log \eta \log \log N)$.

This section proceeds as follows.
We begin with a summary of our algorithm.
We then explain the reasoning underlying the key step (\step{uncompute_sort}) of our algorithm,
which is to reverse a sorting operation on \texttt{target}.
Next we discuss the choice of sorting algorithm, which we require to be a sorting network.
Then, we assess the cost of our algorithm in terms of gate complexity and runtime and we compare this to previous work  in Refs.~\cite{Abrams1997,Abrams1994}.
Finally, we discuss the possibility of antisymmetrizing without sorting and propose an alternative, though more costly, algorithm based on the Fisher-Yates shuffle. Our algorithm consists of the following four steps:

\begin{enumerate}

	\item \label{step:prep_seed}
    \textbf{Prepare \texttt{seed}.}
    Let $f$ be a function chosen so that $f(\eta) \geq \eta^2$ for all $\eta$.
    We prepare an ancillary register called \texttt{seed} in an even superposition of all possible length-$\eta$ strings of the numbers $0,1, \ldots, f(\eta)-1$.
    If $f(\eta)$ is a power of two, preparing \texttt{seed} is easy:
    simply apply a Hadamard gate to each qubit.
    
	\item \label{step:sort_seed}
    \textbf{Sort \texttt{seed}.}
    Apply a reversible sorting network to \texttt{seed}.
    Any sorting network can be made reversible by storing the outcome of each comparator in a second ancillary register called \texttt{record}.
    There are several known sorting networks with polylogarithmic runtime,
    as we discuss below.
    
	\item \label{step:delete_reps}
    \textbf{Delete collisions from \texttt{seed}}.
    As \texttt{seed} was prepared in a superposition of all length-$\eta$ strings,
    it includes strings with repeated entries.
    As we are imposing fermionic exchange symmetry, these repetitions must be deleted.
    We therefore measure \texttt{seed} to determine whether a repetition is present,
    and we accept the result if it is repetition-free.
    We prove in \app{DeleteCollisions} that choosing $f(\eta) \geq \eta^2$ ensures that the probability of success is greater than $1/2$.
    We further prove that the resulting state of \texttt{seed} is disentangled from \texttt{record},
    meaning \texttt{seed} can be discarded after this step.
    
	\item \label{step:uncompute_sort}
    \textbf{Apply the reverse of the sort to \texttt{target}.}
    Using the comparator values stored in \texttt{record},
    we apply each step of the sorting network in reverse order to the sorted array \texttt{target}.
    The resulting state of \texttt{target} is an evenly weighted superposition of each possible permutation of the original values.
    To ensure the correct phase, we apply a controlled-phase gate after each swap.
\end{enumerate}

\step{uncompute_sort} is the key step.
Having prepared (in \step{prep_seed}-\step{delete_reps})
a record of the in-place swaps needed to sort a symmetrized, collision-free array,
we undo each of these swaps in turn on the sorted \texttt{target}.
We employ a sorting network, a restricted type of sorting algorithm,
because sorting networks have comparisons and swaps at a fixed sequence of locations.
By contrast,
many common classical sorting algorithms (like heapsort) choose locations depending on the values in the list.
This results in accessing registers in a superposition of locations in the corresponding quantum algorithm, incurring a linear overhead.
As a result, a quantum heapsort requires $\widetilde{O} \left(\eta^2\right)$ operations, not $\widetilde{O} (\eta)$.
By contrast, no overhead is required for using a fixed sequence of locations.
The implementation of sorting networks in quantum algorithms has previously been considered in Refs.~\cite{Cheng06,Beals13}.

\begin{figure}[b]
\resizebox{.95\linewidth}{!}{
    \includegraphics{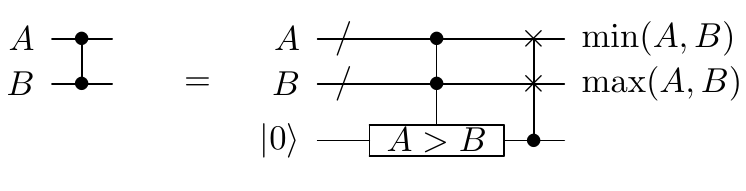}
  }
 \caption{The standard notation for a comparator is indicated on the left. Its implementation as a quantum circuit is shown on the right. In the first step, we compare two inputs with values $A$ and $B$ and save the outcome ($1$ if $A>B$ is true and $0$ otherwise) in a single-qubit ancilla. In the second step, conditioned on the value of the ancilla qubit, the values $A$ and $B$ in the two wires are swapped.} 
 \label{fig:compare}
\end{figure}

Sorting networks are logical circuits that consist of wires 
carrying values and comparator modules applied to pairs
of wires, that compare values and swap them if they are not in the correct order.
Wires represent bit strings (integers are stored in binary) 
in classical sorting networks and qubit strings in their quantum analogues.
A classical comparator is a sort on two numbers, which gives the transformation  $(A,B)\mapsto \left(\min(A,B), \max(A,B)\right)$.
A quantum comparator is its reversible version where we record whether the items were already sorted (ancilla state $\ket{0}$) or the comparator needed to apply a swap (ancilla state $\ket{1}$); see \fig{compare}.

The positions of comparators are set as a predetermined fixed sequence in advance
and therefore cannot depend on the inputs.
This makes sorting networks viable candidates for quantum computing. Many of the sorting networks are also highly parallelizable, thus allowing low-depth, often polylogarithmic, performance. 

\begin{figure}[t]
\centering
\resizebox{.85\linewidth}{!}{
    \includegraphics{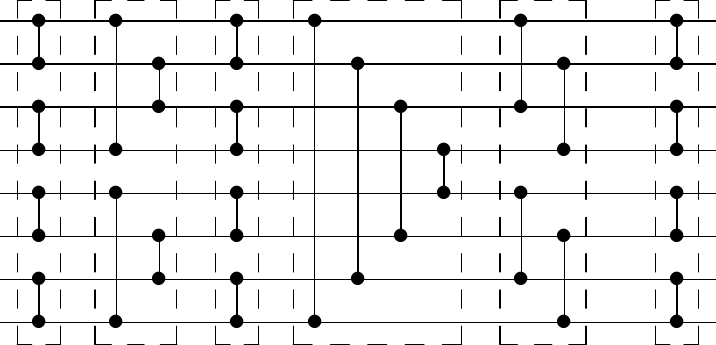}
  }
\caption{Example of a {\em bitonic sort} on 8 inputs. The ancillae necessary to record the results as part of implementing each of the  comparators are omitted for clarity. Comparators in each dashed box can be applied in parallel for depth reduction.}
 \label{fig:bitonic}
\end{figure}

Our algorithm allows for any choice of sorting network.
Several common sort algorithms such as the insert and bubble sorts can be represented as sorting networks.
However, these algorithms have poor time complexity even after parallelization.
More efficient runtime can be achieved, for example, using the bitonic sort \cite{Bat68,Bat93}, which is illustrated for $8$ inputs in \fig{bitonic}. The bitonic sort uses $O(\eta \log^2 \eta)$ comparators and $O(\log^2 \eta)$ depth, thus achieving an exponential improvement in depth compared to common sorting techniques.

Slightly better performance can be obtained using an odd-even mergesort \cite{Bat68}.
The asymptotically best sorting networks have depth $O(\log\eta)$ and complexity $O(\eta\log\eta)$, though there is a large constant which means they are less efficient for realistic $\eta$ \cite{AKS83,Pat90}.
There is also a sorting network with $O(\eta\log\eta)$ complexity with a better multiplicative constant \cite{Good14}, though its depth is $O(\eta\log\eta)$ (so it is not logarithmic).

Assuming we use an asymptotically optimal sorting network,
the circuit depth for our algorithm is $O(\log \eta \log \log N)$
and the gate complexity is $O(\eta \log \eta \log N)$.
The dominant cost of the algorithm comes from \step{sort_seed} and \step{uncompute_sort},
each of which have $O(\eta \log \eta)$ comparators that can be parallelized to ensure the sorting network executes only $O(\log \eta)$ comparator rounds.
Each comparator for \step{uncompute_sort} has a complexity of $O(\log N)$ and a depth of $O(\log \log N)$, as we show in \app{QuantumComparators}.
The comparators for \step{sort_seed} have complexity $O(\log \eta)$ and depth $O(\log \log \eta)$, which is less because $\eta<N$.
Thus \step{sort_seed} and \step{uncompute_sort} each have gate complexity $O(\eta \log \eta \log N)$ and runtime $O(\log \eta \log \log N)$.

The other two steps in our algorithm have smaller cost.
\step{prep_seed} has constant depth and $O(\eta\log\eta)$ complexity.
\step{delete_reps} requires $O(\eta)$ comparisons
because only nearest-neighbor comparisons need be carried out on \texttt{seed} after sorting.
These comparisons can be parallelized over two rounds, with complexity $O(\eta \log \eta)$ and
circuit depth $O(\log\log \eta)$.
Then the result for any of the registers being equal is computed in a single qubit, which has complexity $O(\eta)$ and depth $O(\log\eta)$.
Thus the complexity of \step{delete_reps} is $O(\eta \log \eta)$ and
the total circuit depth is $O(\log \eta)$.
We give further details  in \app{QuantumComparators}. Thus, our algorithm has an exponential runtime improvement over  the proposal in Refs.~\cite{Abrams1997,Abrams1994}.
We also have a quadratic improvement in gate complexity,
which is $\widetilde{O}(\eta)$ for our algorithm but $\widetilde{O} (\eta^2)$ for Refs.~\cite{Abrams1997,Abrams1994}.

Our runtime is likely optimal for symmetrization, at least in terms of the $\eta$ scaling.
Symmetrization takes a single computational basis state and generates a superposition of $\eta!$ computational basis states.
Each single-qubit operation can increase the number of states in the superposition by at most a factor of two, and two-qubit operations can increase the number of states in the superposition by at most a factor of four.
Thus, the number of one- and two-qubit operations is at least
$\log_2 (\eta!) = O(\eta\log\eta)$.
In our algorithm we need this number of operations between the registers.
If that is true in general, then $\eta$ operations can be parallelized, resulting in minimum depth $O(\log\eta)$.
It is more easily seen that the total number of registers used is optimal.
There are $O(\eta\log\eta)$ ancilla qubits due to the number of steps in the sort, but the number of qubits for the system state we wish to symmetrize is $O(\eta\log N)$, which is asymptotically larger.

Our quoted asymptotic runtime and gate complexity scalings assume the use of sorting networks that are asymptotically optimal.
However, these algorithms have a large constant overhead making it more practical to use an odd-even mergesort, leading to depth $O(\log^2\eta \log\log N)$. Note that is possible to obtain complexity $O(\eta\log\eta\log N)$ and number of ancilla qubits $O(\eta\log\eta)$ with a better scaling constant using the sorting network of Ref.~\cite{Good14}.

Given that the cost of our algorithm is dictated by the cost of sorting algorithms,
it is natural to ask if it is possible to antisymmetrize without sorting.
Though the complexity and runtime both turn out to be significantly worse than our sort-based approach,
we suggest an alternative antisymmetrization algorithm based on the Fisher-Yates shuffle.
The Fisher-Yates shuffle is a method for applying to a length-$\eta$ target array a permutation chosen uniformly at random using a number of operations scaling as $O(\eta)$.
Our algorithm indexes the positions to be swapped,
thereby increasing the complexity to $\widetilde{O}(\eta^2)$.
Briefly put, our algorithm generates a superposition of states as in Step II of Ref.~\cite{Abrams1997},
then uses these as control registers to apply the Fisher-Yates shuffle to the orbital numbers.
The complexity is $O(\eta^2\log N)$, with a factor of $\log N$ due to the size of the registers.
We reset the control registers, thereby disentangling them, using $O(\eta\log\eta)$ ancillae.
We provide more details of this approach in \app{FisherYatesShuffle}.

To conclude this section,
we have presented an algorithm for antisymmetrizing a sorted, repetition-free quantum register.
The dominant cost of our algorithm derives from the choice of sorting network,
whose asymptotically optimal gate count complexity and runtime are, respectively, $O (\eta \log \eta \log N)$ and  $O (\log \eta \log \log N)$.
This constitutes a polynomial improvement in the first case and exponential in the second case over previous work  in Refs.~\cite{Abrams1997,Abrams1994}.
As in Ref.~\cite{Abrams1997}, our antisymmetrization algorithm constitutes a key step for preparing fermionic wavefunctions in first quantization.

\section{Fewer Phase Estimation Repetitions by Partial Eigenstate Projection Rejection}
\label{sec:rejection}

Once the initial state $\ket{\varphi}$ has been prepared, it typically will not be exactly the ground state (or other eigenstate) desired.
In the usual approach, one would perform phase estimation repeatedly, in order to obtain the desired eigenstate $\ket{k}$.
The number of repetitions needed scales inversely in $\alpha_k = | \braket{\varphi}{k} |^2$, increasing the complexity.
We propose a practical strategy for reducing this cost which is particularly relevant for quantum chemistry. Our approach applies if one seeks to prepare the ground state with knowledge of an upper bound on the ground state energy $\tilde{E}_0$, together with the promise that $E_0 \leq \tilde{E}_0 < E_1$. With such bounds available, one can reduce costs by restarting the phase estimation procedure as soon as the energy is estimated to be above $\tilde{E}_0$ with high probability.
That is, one can perform a phase estimation procedure that gradually provides estimates of the phase to greater and greater accuracy, for example as in Ref.~\cite{HigginsNAT07}.
If at any stage the phase is estimated to be above $\tilde{E}_0$ with high probability, then the initial state can be discarded and re-prepared.

Performing phase estimation within error $\epsilon$ typically requires evolution time for the Hamiltonian of $1/\epsilon$, leading to complexity scaling as $1/\epsilon$.
This means that, if the state is the first excited state, then an estimation error less than $E_1-\tilde{E}_0$ will be sufficient to show that the state is not the ground state.
The complexity needed would then scale as $1/(E_1-\tilde{E}_0)$.
In many cases, the final error required, $\epsilon_f$, will be considerably less than $E_1-\tilde{E}_0$, so the majority of the contribution to the complexity comes from measuring the phase with full precision, rather than just rejecting the state as not the ground state.

Given the initial state $\ket{\varphi}$ which has initial overlap of $\alpha_0$ with the ground state, if we restart every time the energy is found to be above $\tilde{E}_0$, then the contribution to the complexity is $1/[\alpha_0(E_1-\tilde{E}_0)]$.
There will be an additional contribution to the complexity of $1/\epsilon_f$ to obtain the estimate of the ground state energy with the desired accuracy, giving an overall scaling of the complexity of
\begin{equation}
O\left(\frac 1{\alpha_0(E_1-\tilde{E}_0)} + \frac 1{\epsilon_f}\right).
\end{equation}
In contrast, if one were to perform the phase estimation with full accuracy every time, then the scaling of the complexity would be $O(1/(\alpha_0 \epsilon_f))$. Provided $\alpha_0(E_1-\tilde{E}_0)>\epsilon_f$, the method we propose would essentially eliminate the overhead from $\alpha_0$.

In cases where $\alpha_0$ is very small, it would be helpful to apply amplitude amplification.
A complication with amplitude amplification is that we would need to choose a particular initial accuracy to perform the estimation.
If a lower bound on the excitation energy, $\tilde{E}_1$, is known, then we can choose the initial accuracy to be $\tilde{E}_1-\tilde{E}_0$.
The success case would then correspond to not finding that the energy is above $\tilde{E}_0$ after performing phase estimation with that precision.
Then amplitude amplification can be performed in the usual way, and the overhead for the complexity is $1/\sqrt{\alpha_0}$ instead of $1/\alpha_0$.

All of this discussion is predicated on the assumption that there are cases where $\alpha_0$ is small enough to warrant using phase estimation as part of the state preparation process and where a bound meeting the promises of $\tilde{E}_0$ is readily available. We now discuss why these conditions are anticipated for many problems in quantum chemistry. Most chemistry is understood in terms of mean-field models (e.g.\ molecular orbital theory, ligand field theory, the periodic table, etc.). Thus, the usual assumption (empirically confirmed for many smaller systems) is that the ground state has reasonable support on the Hartree-Fock state (the typical choice for $\ket{\varphi}$) \cite{Wang2008,Veis2014,McClean2014,BabbushTrotter}. However, this overlap will decrease as a function of both basis size and system size. As a simple example, consider a large system composed of $n$ copies of non-interacting subsystems. If the Hartree-Fock solution for the subsystem has overlap $\alpha_0$, then the Hartree-Fock solution for the larger system has overlap of exactly $\alpha_0^n$, which is exponentially small in $n$.

It is literally plain-to-see that the electronic ground state of molecules is often protected by a large gap. The color of many molecules and materials is the signature of an electronic excitation from the ground state to first excited state upon absorption of a photon in the visible range (around 0.7 Hartree); many clear organics have even larger gaps in the UV spectrum. Visible spectrum $E_1 - E_0$ gaps are roughly a hundred times larger than the typical target accuracy of $\epsilon_f = 0.0016$ Hartree (``chemical accuracy'')\footnote{The rates of chemical reactions are proportional to $e^{- \beta \Delta A} / \beta$ where $\beta$ is inverse temperature and $\Delta A$ is a difference in free energy between reactants and the transition state separating reactants and products. Chemical accuracy is defined as the maximum error allowable in $\Delta A$ such that errors in the rate are smaller than a factor of ten at room temperature \cite{Aspuru-Guzik2005}.}. Furthermore, in many cases the first excited state is perfectly orthogonal to the Hartree-Fock state for symmetry reasons (e.g.\ due to the ground state being a spin singlet and the excited state being a spin triplet). Thus, the gap of interest is really $E^* - E_0$ where $E^* = \min_{k>0} E_k$ subject to $|\braket{\varphi}{k}|^2 > 0$. Often the $E^* - E_0$ gap is much larger than the $E_1 - E_0$ gap.

For most problems in quantum chemistry a variety of scalable classical methods are accurate enough to compute upper bounds on the ground state energy $\tilde{E}_0$ such that $E_0 \leq \tilde{E}_0 < E^*$, but not accurate enough to obtain chemical accuracy (which would require quantum computers). Classical methods usually produce upper bounds when based on the variational principle. Examples include mean-field and Configuration Interaction Singles and Doubles (CISD) methods \cite{Helgaker2002}.

As a concrete example, consider a calculation on the water molecule in its equilibrium geometry (bond angle of $104.5^\circ$, bond length of $0.9584$ \AA) in the minimal (STO-3G) basis set performed using OpenFermion \cite{openfermion} and Psi4 \cite{Psi4}. For this system, $E_0 = -75.0104$ Hartree and $E_1 = -74.6836$ Hartree. However, $\braket{\varphi}{1}$ = 0 and $E^* = -74.3688$ Hartree. The classical mean-field energy provides an upper bound on the ground state energy of $\tilde{E}_0 = -74.9579$ Hartree.
Therefore $E^*-\tilde{E}_0\approx 0.6$ Hartree, which is about $370$ times $\epsilon_f$.
Thus, using our strategy, for $\alpha_0 > 0.003$ there is very little overhead due to the initial state $\ket{\varphi}$ not being the exact ground state. In the most extreme case for this example, that represents a speedup by a factor of more than two orders of magnitude. However, in some cases the ground state overlap might be high enough that this technique provides only a modest advantage. While the Hartree-Fock state overlap in this small basis example is $\alpha_0 = 0.972$, as the system size and basis size grow we expect this overlap will decrease (as argued earlier).

Another way to cause the overlap to decrease is to deviate from equilibrium geometries \cite{Wang2008,Veis2014}. For example, we consider this same system (water in the minimal basis) when we stretch the bond lengths to $2.25\times$ their normal lengths.
In this case, $E_0 = -74.7505$ Hartree, $E^* = -74.6394$ Hartree, and $\alpha_0 = 0.107$. The CISD solution provides an upper bound $\tilde{E}_0 = -74.7248$.
In this case, $E^*-\tilde{E}_0\approx 0.085$ Hartree, about $50$ times $\epsilon_f$.
Since $\alpha_0 > 0.02$, here we speed up state preparation by roughly a factor of $\alpha_0^{-1}$ (more than an order of magnitude).

\section{Phase Estimation Unitaries without Approximation}
\label{sec:no_error}
Normally, the phase estimation would be performed by Hamiltonian simulation.
That introduces two difficulties: first, there is error introduced by the Hamiltonian simulation that needs to be taken into account in bounding the overall error, and second, there can be ambiguities in the phase that require simulation of the Hamiltonian over very short times to eliminate.

These problems can be eliminated if one were to use Hamiltonian simulation via a quantum walk, as in Refs.~\cite{Childs2010b,Berry2012}.
There, steps of a quantum walk can be performed exactly, which have eigenvalues related to the eigenvalues of the Hamiltonian.
Specifically, the eigenvalues are of the form $\pm e^{\pm i\arcsin(E_k/\lambda)}$.
Instead of using Hamiltonian simulation, it is possible to simply perform phase estimation on the steps of that quantum walk, and invert the function to find the eigenvalues of the Hamiltonian.
That eliminates any error due to Hamiltonian simulation.
Moreover, the possible range of eigenvalues of the Hamiltonian is automatically limited, which eliminates the problem with ambiguities.

The quantum walk of Ref.~\cite{Berry2012} does not appear to be appropriate for quantum chemistry, because it requires an efficient method of calculating matrix entries of the Hamiltonian.
That is not available for the Hamiltonians of quantum chemistry, but they can be expressed as sums of unitaries, as for example discussed in Ref.~\cite{BabbushSparse1}.
It turns out that the method called qubitization \cite{Low2016} allows one to take a Hamiltonian given by a sum of unitaries, and construct a new operation with exactly the same functional dependence on the eigenvalues of the Hamiltonian as for the quantum walk in Refs.~\cite{Childs2010b,Berry2012}.

Next, we summarize how qubitization works \cite{Low2016}.
One assumes black-box access to a signal oracle $V$ that encodes $H$ in the form:
 \begin{align}
 \left(\proj{0}_a \otimes \openone_s\right) V \left(\proj{0}_a \otimes \openone_s\right) = \proj{0}_a \otimes H / \lambda
 \end{align}
 where $\ket{0}_a$ is in general a multi-qubit ancilla state in the computational basis, $\openone_s$ is the identity gate on the system register and $\lambda \ge \|H\|$ is a normalization constant.
For Hamiltonians given by a sum of unitaries,
\begin{equation}
\label{eq:lcu}
H=\sum^{d-1}_{j=0}a_jU_j \quad \quad a_j > 0,
\end{equation}
one constructs
\begin{equation}
U = (A^\dag\otimes \openone)\operatorname{SELECT-U}(A\otimes \openone),
\end{equation}
where $A$ is an operator for state preparation acting as
\begin{equation}
A\ket{0}=\sum^{d-1}_{j=0}\sqrt{a_j/\lambda}\ket{j}
\end{equation}
with $\lambda = \sum^{d-1}_{j=0}a_j$,
and
\begin{equation}
\operatorname{SELECT-U}=\sum^{d-1}_{j=0}\proj{j}\otimes U_j.
\end{equation}

For $U$ that is Hermitian, we can simply take $V=U$. This is the case for any local Hamiltonian that can be written as a weighted sum of tensor products of Pauli operators, since tensor products of Pauli operators are both unitary and Hermitian. More general strategies for representing Hamiltonians as linear combinations of unitaries, as in \eq{lcu}, are discussed in \cite{Berry2013}, related to the sparse decompositions first described in \cite{Aharonov2003}.  If $U$ is not Hermitian, then we may construct a Hermitian $V$ as
\begin{equation}
V=\ket{+}\!\!\bra{-}\otimes U+\ket{-}\!\!\bra{+}\otimes U^\dag
\end{equation}
where $\ket{\pm}=\frac{1}{\sqrt{2}}(\ket{0}\pm\ket{1})$.
The multiqubit ancilla labelled ``$a$'' would then include this additional qubit, as well as the ancilla used for the control for $\operatorname{SELECT-U}$.
In either case we can then construct a unitary operator called the qubiterate as follows:
\begin{align}
 W&=i(2\proj{0}_a\otimes \openone_s - \openone)V.
\end{align}

The qubiterate transforms each eigenstate $\ket{k}$ of $H$ as
\begin{align}
W \ket{0}_a \ket{k}_s &= i\frac{E_k}{\lambda} \ket{0}_a \ket{k}_s + i\sqrt{1-\left|\frac{E_k}{\lambda}\right|^2}\ket{0k^\perp}_{as}\\
W\ket{0k^\perp}_{as} &= i\frac{E_k}{\lambda} \ket{0k^\perp}_{as} - i\sqrt{1-\left|\frac{E_k}{\lambda}\right|^2}\ket{0}_a\ket{k}_{s}
 \end{align} 
 where $\ket{0k^\perp}_{as}$ has no support on $\ket{0}_a$. Thus, $W$ performs rotation between two orthogonal states $\ket{0}_a\ket{k}_s$ and $\ket{0k^\perp}_{as}$. Restricted to this subspace, the qubiterate may be diagonalized as
 \begin{align}
  \label{eq:qubiterate_diagonal}
W\ket{\pm k}_{as} &= \mp e^{\mp i \arcsin(E_k/\lambda)}\ket{\pm k}_{as}
\\
\ket{\pm k}_{as} &= \frac{1}{\sqrt{2}}\left(\ket{0}_a\ket{k}_s \pm \ket{0k^\perp}_{as}\right).
 \end{align}
This spectrum is exact, and identical to that for the quantum walk in Refs.~\cite{Childs2010b,Berry2012}. This procedure is also simple, requiring only two queries to $U$ and a number of gates to implement the controlled-$Z$ operator $(2\proj{0}_a\otimes \openone_s-\openone)$ scaling linearly in the number of controls.
 
We may replace the time evolution operator with the qubiterate $W$ in phase estimation, and phase estimation will provide an estimate of $\arcsin(E_k/\lambda)$ or $\pi-\arcsin{(E_k/\lambda)}$.
In either case taking the sine gives an estimate of $E_k/\lambda$, so it is not necessary to distinguish the cases.
Any problems with phase ambiguity are eliminated, because performing the sine of the estimated phase of $W$ yields an unambiguous estimate for $E_k$.
Note also that $\lambda\ge\|H\|$ implies that $|E_k/\lambda|\le 1$.

More generally, any unitary operation $e^{i f(H)}$ that has eigenvalues related to those of the Hamiltonian would work so long as the function $f(\cdot): \mathbb{R} \rightarrow (-\pi,\pi)$ is known in advance and invertible. One may perform phase estimation to obtain a classical estimate of ${f}(E_{k})$, then invert the function to estimate $E_k$. To first order, the error of the estimate would then propagate like \begin{equation}
\sigma_{E_{k}} = \left| \left(\left.\frac{d f}{dx}\right|_{x=E_{k}} \right)\right|^{-1} \!\! \sigma_{f\left(E_{k}\right)}.
 \end{equation}

In our example, with standard deviation $\sigma_\text{phase}$ in the phase estimate of $W$, the error in the estimate is
\begin{align}
\sigma_{E_{k}} = \sigma_\text{phase} \sqrt{\lambda^2-E_k^2} \le \lambda \,\sigma_\text{phase} \, .
\end{align}
Obtaining uncertainty $\epsilon$ for the phase of $W$ requires applying $W$ a number of times scaling as $1/\epsilon$.
Hence, obtaining uncertainty $\epsilon$ for $E_k$ requires applying $W$ a number of times scaling as $\lambda/\epsilon$.
For Hamiltonians given by sums of unitaries, as in chemistry, each application of $W$ uses $O(1)$ applications of state preparations and $\operatorname{SELECT-U}$ operations.
In terms of these operations, the complexities of       \sec{rejection} have multiplying factors of $\lambda$.

\section{Conclusion}

We have described three techniques which we expect will be practical and useful for the quantum simulation of fermionic systems. Our first technique provides an exponentially faster method for antisymmetrizing configuration states, a necessary step for simulating fermions in first quantization. We expect that in virtually all circumstances the gate complexity of this algorithm will be nearly trivial compared to the cost of the subsequent phase estimation. Then, we showed that when one has knowledge of an upper bound on the ground state energy that is separated from the first excited state energy, one can prepare ground states using phase estimation with lower cost. We discussed why this situation is anticipated for many problems in chemistry and provided numerics for a situation in which this trick reduced the gate complexity of preparing the ground state of molecular water by more than an order of magnitude. Finally, we explained how qubitization \cite{Low2016} provides a unitary that can be used for phase estimation without introducing the additional error inherent in Hamiltonian simulation.

We expect that these techniques will be useful in a variety of contexts within quantum simulation. In particular, we anticipate that the combination of the three techniques will enable exceptionally efficient quantum simulations of chemistry based on methods similar to those proposed in \cite{Kivlichan2016}. While specific gate counts will be the subject of a future work, we conjecture that such techniques will enable simulations of systems with roughly a hundred electrons on a million point grid with fewer than a billion T gates. With such low T counts, simulations such as the mechanism of Nitrogen fixation by ferredoxin, explored for quantum simulation in \cite{Reiher2017}, should be practical to implement within the surface code in a reasonable amount of time with fewer than a few million physical qubits and error rates just beyond threshold.  This statement is based on the time and space complexity of magic state distillation (usually the bottleneck for the surface code) estimated for superconducting qubit architectures in \cite{Fowler2012}, and in particular, the assumption that with a billion T gates or fewer one can reasonably perform state distillation in series using only a single T factory.

\section*{Acknowledgements}

The authors thank Matthias Troyer for relaying the idea of Alexei Kitaev that phase estimation could be performed without Hamiltonian simulation. We thank Jarrod McClean for discussions about molecular excited state gaps. We note that arguments relating the colors of compounds to their molecular excited state gaps in the context of quantum computing have been popular for many years, likely due to comments of Al\'{a}n Aspuru-Guzik.  DWB is funded by an Australian Research Council Discovery Project (Grant No.\ DP160102426).

\section*{Author Contributions}
DWB proposed the algorithms of \sec{symm} and the basic idea behind \sec{rejection} as solutions to issues raised by RB. MK, AS and YRS worked out and wrote up the details of \sec{symm} and associated appendices. RB connected developments to chemistry simulation, conducted numerics, and wrote \sec{rejection} with input from DWB. Based on discussions with NW, GHL suggested the basic idea of \sec{no_error}. CG helped to improve the gate complexity of our comparator circuits. Remaining aspects of the paper were written by RB and DWB with assistance from MK, AS and YRS.


\bibliographystyle{apsrev4-1}
\bibliography{references,Mendeley}

\appendix

\section{Analysis of `Delete Collisions' Step}
\label{app:DeleteCollisions}
In this appendix, we explain the most difficult-to-understand step of our algorithm:
the step in which we delete collisions from \texttt{seed}.
There are two important points that require explanation.
First, we have to show that the probability of failure is small.
Second, we have to show that the resulting state of \texttt{seed} is disentangled from \texttt{record},
as we wish to uncompute \texttt{record} during the final step of our algorithm.

To explain these two points, we begin with an analysis of the state of \texttt{seed} after Step 1.
The state of \texttt{seed} is
\begin{equation}
	\frac{1}{f(\eta)^{\eta/2}}
    \sum_{\ell_0, \ldots, \ell_{\eta-1}=0}^{f(\eta)-1}
    \ket{\ell_0, \ldots, \ell_{\eta-1}}.
\end{equation}
We can decompose the state space of \texttt{seed} into two orthogonal subspaces:
the `repetition-free' subspace
\begin{equation}
	\operatorname{span} \left\{ \ket{\ell_0, \ldots, \ell_{\eta-1}} | \forall i \neq j: \ell_i \neq \ell_j \right\}
\end{equation}
and its orthogonal complement.
If we project the state of \texttt{seed} onto the repetition-free subspace,
we obtain the unnormalized vector
\begin{equation}
	\frac{1}{f(\eta)^{\eta/2}}
    \sum_{0 \leq \ell_0 < \ldots < \ell_{\eta-1}< f(\eta)}
    \sum_{\sigma \in S_\eta}
    \ket{ \sigma \left(\ell_0, \ldots, \ell_{\eta-1}\right) }.
\end{equation}
The norm of this vector is
\begin{equation}
	\frac{\eta!}{f(\eta)^\eta}
    \begin{pmatrix} f(\eta) \\ \eta \end{pmatrix},
\end{equation}
which is equal to $1-C(f(\eta), \eta)$ in the terminology of Proposition~A.1 in~\cite{bella}.

We sort the register in Step 2 before detecting repetitions in Step 3, because then it is only necessary to check adjacent registers.
The probability of repetitions in unaffected by the sort, because it is unitary and does not affect whether there are repetitions.
Therefore the probability of failure (detection of a repetition) in Step 3 is equal to $C(f(\eta), \eta)$.
Using Proposition~A.1 in~\cite{bella},
the probability of failure is bounded as
\begin{equation}
	\operatorname{Pr} (\text{repetition})
    =		C(f(\eta), \eta)
    \leq	\frac{\eta(\eta-1)}{2f(\eta)},
\end{equation}
which is less than $1/2$ for $f(\eta) \geq \eta^2$.
The repetition-free outcome can therefore be achieved after fewer than two attempts on average.
One can improve the success probability by using a larger function $f$ or by using amplitude amplification.

We now show that $\texttt{seed}\otimes\texttt{record}$ is in an unentangled state after Step 3.
After Step 1,
the state of $\texttt{seed}\otimes\texttt{record}$ projected to the repetition-free subspace can be represented (up to normalization) as
\begin{equation}
    \sum_{0 \leq \ell_0 < \ldots < \ell_{\eta-1}< f(\eta)}
    \sum_{\sigma \in S_\eta}
    \ket{\sigma \left(\ell_0, \ldots, \ell_{\eta-1}\right)}_\texttt{seed}
    \ket{\iota}_\texttt{record}.
\end{equation}
Here we represent the state of \texttt{record} as a recording of all permutations we have applied to \texttt{seed};
$\iota$ represents the identity permutation.
During Step 2, a sequence of permutations $\sigma_1, \ldots, \sigma_T$ (where $T$ depends on the choice of sorting network) is applied to \texttt{seed} and recorded on \texttt{record}.
This sequence of permutations is chosen so that
\begin{equation}
\label{eq:sort_cond_1}
	\sigma_T \circ \cdots \circ \sigma_1 \circ \sigma
    \left(\ell_0, \ldots, \ell_{\eta-1} \right)
    =	\left(\ell_0, \ldots, \ell_{\eta-1} \right),
\end{equation}
where $0 \leq \ell_0 < \ldots < \ell_{\eta-1} < f(\eta)$.
That is to say,%
\footnote{%
Note that no condition like \eq{sort_cond_2} holds in the orthogonal complement of the repetition-free subspace.
There are multiple permutations that sort an unsorted array that has repeated elements,
so the choice of $\sigma$ would be ambiguous.}%
\begin{equation}
\label{eq:sort_cond_2}
	\sigma_T \circ \cdots \circ \sigma_1 \circ \sigma
    =	\iota.
\end{equation}
Therefore, the state of $\texttt{seed}\otimes\texttt{record}$ after Step 3 is (up to normalization)
\begin{equation}
    \sum_{0 \leq \ell_0 < \ldots < \ell_{\eta-1}< f(\eta)}\!\!\!
    \ket{\ell_0, \ldots, \ell_{\eta-1}}_\texttt{seed}
    \sum_{\sigma \in S_\eta}
    \ket{\sigma_1, \ldots, \sigma_T}_\texttt{record}.
\end{equation}
This is a product state.
Therefore, \texttt{seed} can be discarded after Step 3 without affecting \texttt{record}.

\section{Quantum Sorting}
\label{app:QuantumComparators}

\subsection{Resource Analysis of Quantum Sorting Networks}

In this section we expand on the resource analysis of implementing quantum sorting networks. We also illustrate that for small number of inputs to be sorted (up to $\eta=20$), concrete bounds have been 
derived for optimized circuit depth as well as the 
number of comparators. This may be of interest and 
useful for implementing quantum simulations of small molecules, 
also in view of the observation that $\eta\approx 20$ 
is nearly reaching a number of electrons for where classical simulations become intractable.

Optimizing sorting networks for small inputs is an active research area in parallel programming.  Knuth \cite{knuth1997art} and later Codish \textit{et al.}~\cite{CODISH2016} gave networks for sorting up to 17 numbers that were later shown to be optimal in depth, 
and up to $\eta \leq 10$ also optimal in the number of comparators. 
Optimizations for up to 20 inputs have recently been achieved, 
see Table~1 in~\cite{CODISH2016}. 
In such optimizations one typically distinguishes between the 
optimal depth problem 
and the problem of minimizing the overall number of 
comparators. 
For illustration, the best known sorting networks for 20 numbers require depth 11 and 92 comparators, with lower bounds reported as 10 and 73 respectively. Efficient sorting networks can be produced by in-place merging of sorting networks with smaller sizes.  However, this procedure necessarily produces some overhead.

For our resource analysis we assume that the quantum sorting network 
has $\eta$ wires, where each wire represents a quantum register of length $d$ 
(i.e., consists of $d$ qubits). The resource requirement for implementing 
the quantum sort is obtained by taking 
the (classical) sorting network depth or the overall number of comparators involved and multiplying 
it by the corresponding resources needed to construct a comparator. As explained above, the latter 
requires one query to a comparison oracle, whose 
circuit implementation and complexity are provided in \app{ComparisonOracle}, 
and a conditional swap applied to the compared registers of size $d$
controlled by the single-qubit ancilla holding the result of the comparison. 
 
The construction of the  comparison oracle as well as the implementation of the conditional swaps both yield a network consisting predominantly of Toffoli, \textsc{Not} and 
\textsc{CNot} gates 
requiring $O(d)$ elementary gate operations but only $O(\log d)$ circuit depth. Indeed, 
as shown in \app{ComparisonOracle}, the comparison oracle can be implemented such that 
the operations can mostly be performed in parallel with only $O(\log d)$ circuit depth. 

When implementing conditional swaps on two registers of size $d$  as part of a comparator, 
all elementary swaps between the corresponding qubits of
these registers must be controlled by the very same ancilla qubit,
namely the one encoding the result of the comparison oracle. 
This suggests having to perform all the controlled swaps in sequence, 
as they all are to be controlled by the same qubit, which would imply  depth scaling $O(d)$ 
rather than $O(\log d)$. Yet the conditional swaps can also
be parallelized. This can be achieved by first copying the 
bit of the ancilla holding the result of the comparison 
to $d-1$ additional ancillae, all initialized in $\ket{0}$. 
Such an expansion of the result to $d$ copies can 
be attained with a parallelized arrangement 
of $O(d)$ $\textsc{CNot}s$ but with circuit depth 
only $O(\log d)$. After copying, all the $d$ 
controlled elementary swaps can then be executed in parallel   
(by using the additional ancillae) with circuit depth only $O(1)$. 
After executing the swaps, the $d-1$ additional ancillae used for 
holding the copied result of comparison are uncomputed again, 
by reversing the copying process. 
While this procedure requires  $O(d)$ ancillary space overhead, it optimizes 
the depth. The overall space overhead of the quantum comparator is also $O(d)$. 

Taking $d=\lceil\log N\rceil$ (the largest registers used in Step 4 of our sort-based 
antisymmetrization algorithm), conducting the quantum bitonic sort, for instance, 
thus requires $O(\eta \log^2 (\eta) \log N)$ elementary gates but only 
$O(\log^2(\eta) \log\log N)$ circuit depth, while the overall worst-case 
ancillary space overhead amounts to $O(\eta \log^2 (\eta) \log N) $.

\subsection{Comparison Oracle}
\label{app:ComparisonOracle}

Here we describe how to implement reversibly the comparison of the value held in 
one register with the value carried by a second equally-sized register, and store the 
result (larger or not) in a single-qubit ancilla. 
We term the corresponding unitary process a {\em \lq{}comparison oracle\rq{}}. 
We need to use it for implementing the comparator modules 
of quantum sorting networks as well as in our antisymmetrization 
approach based on the quantum Fisher-Yates shuffle. We first explain a naive method for comparison with depth linear in the length of the involved registers. 
In the second step we then convert this prototype into an 
algorithm with depth logarithmic in the register length 
using a divide and conquer approach.

Let $\texttt{A}$ and $\texttt{B}$ denote the two equally sized registers to be compared, 
and $A$ and $B$ the values held by these two registers. 
To determine whether $A>B$ or $A<B$ or $A=B$, 
we compare the registers 
in a {\em bit-by-bit} fashion, 
starting with their most significant 
bits and going down to 
their least significant bits. 
At the very first occurrence  of an $i$ such that 
 $\texttt{A}[i]\not=\texttt{B}[i]$, i.e., either $\texttt{A}[i]=1$ and $\texttt{B}[i]=0$ or $\texttt{A}[i]=0$ and $\texttt{B}[i]=1$, we know that $A>B$ in the first case and $A<B$ in the second case. If $\texttt{A}[i]=\texttt{B}[i]$ for all $i$, 
then $A=B$. We now show how to infer and record the result in a reversible way.

To achieve a reversible comparison, we employ two ancillary registers, each consisting of $d$ qubits, and each initialized to state $\ket{0}^{\otimes d}$, respectively.
We denote them by $\texttt{A}\rq{}$ and $\texttt{B}\rq{}$. 
They are introduced for the purpose of recording the result 
of bitwise comparison as follows. 
$\texttt{A}\rq{}[i]=1$ implies that after $i$ bitwise comparisons 
we know with certainty that $A=\text{max}(A,B)$, while   
$\texttt{B}\rq{}[i]=1$ implies $B=\text{max}(A,B)$. 
These implications can be achieved 
by the following protocol, which is illustrated 
by a simple example in \tab{comparison-table-example}.

To start, at $i=0$ we compare the most significant bits $\texttt{A}[0]$ and $\texttt{B}[0]$, and
write $1$ into ancilla $\texttt{A}\rq{}[0]$ if $\texttt{A}[0]>\texttt{B}[0]$,
or write $1$ into ancilla $\texttt{B}\rq{}[0]$ if $\texttt{A}[0]<\texttt{B}[0]$.
Otherwise the ancillas remain as $0$.
For each $i>0$,
if $\texttt{A}\rq{}[i-1]=0$ and $\texttt{B}\rq{}[i-1]=0$ we compare $\texttt{A}[i]$ and $\texttt{B}[i]$ and record the outcome to $\texttt{A}\rq{}[i]$ and $\texttt{B}\rq{}[i]$ in the same way as for $i=0$. If however 
$\texttt{A}\rq{}[i-1]=1$ and $\texttt{B}\rq{}[i-1]=0$, we already know that $A>B$, so we set $\texttt{A}\rq{}[i]=1$ and $\texttt{B}\rq{}[i]=0$. Similarly,  $\texttt{A}\rq{}[i-1]=0$ and $\texttt{B}\rq{}[i-1]=1$ 
implies $A<B$, so we set $\texttt{A}\rq{}[i]=0$ and $\texttt{B}\rq{}[i]=1$. 
We continue doing so until we reach the least significant bits. This results in the least significant bits of the ancillary registers  $\texttt{A}\rq{}$ and $\texttt{B}\rq{}$ holding information about $\text{max}(A,B)$. 
If these least significant bits are both 0, 
then $A=B$.
At the end the least significant bit of $\texttt{A}\rq{}$ has value $1$ if $A>B$, and $0$ if $A\le B$.
This bit can be copied to an \texttt{output} register, and the initial sequence of operations reversed to erase the other ancilla qubits.

\begin{table}[tb]
\centering
\begin{tabular}{||c||c|c|c|c|c|c|c|c|c||} \hline\hline
Register  & $i$=0 & $i$=1 & $i$=2 & $i$=3  & $i$=4  & $i$=5  & $i$=6  & $i$=7  & $i$=8   \\  \hline\hline
\texttt{A}  & 0 & 0& 0 & 0 & 1 &0 &1 &0&1\\ \hline
\texttt{B}  & 0 & 0& 0 & 0 & 0 &1 &1 &1&0\\ \hline
$\texttt{A}\rq{}$    &0 & 0& 0 & 0 & 1 &1 &1 &1&1\\ \hline
$\texttt{B}\rq{}$   & 0 &0 & 0 & 0 & 0 &0 &0 &0&0\\ \hline\hline
\end{tabular}
\caption{Example illustrating the idea of reversible bitwise comparison. 
Here, $d=9$, the value held in register $\texttt{A}$ is 21 and the value held in register $\texttt{B}$ is 14.  The index $i$ labels the bits of the registers, with $i=0$ designating the most significant bits, respectively. Observe that the first occurrence  of $\texttt{A}[i]\not=\texttt{B}[i]$ is for $i=4$, at which stage the value of ancilla $\texttt{A}\rq{}[4]$ is switched to $1$, as $\texttt{A}[4]>\texttt{B}[4]$. This change causes all lesser significant bits of $\texttt{A}\rq{}$ 
also to be switched to 1, whereas all bits of $\texttt{B}\rq{}$ remain 0. Thus, the least significant bits of $\texttt{A}\rq{}$ and $\texttt{B}\rq{}$ contain information about which number is larger. Here, $\texttt{A}\rq{}[8]=1$ implies $A>B$.}
\label{tab:comparison-table-example}
\end{table}

While this algorithm works, it has the drawback 
that the bitwise comparison is conducted {\em sequentially}, which results in 
circuit-depth scaling $O(d)$. It also uses more ancilla qubits than necessary. We can improve upon this. 
We can reduce the number of ancilla qubits by reusing some input bits as output bits, and we can achieve a depth scaling of $O(\log d)$ 
by parallelizing the bitwise comparison. 
To introduce a parallelization, observe the following. 
Let us split the register \texttt{A} into two parts: $\texttt{A}_1$ consisting of the first approximately $d/2$ bits and $\texttt{A}_2$ consisting of the remaining approximately $d/2$ bits. Split register $\texttt{B}$ in the very same way into subregisters $\texttt{B}_1$ and $\texttt{B}_2$. We can then determine which number is larger (or whether both are equal) for each pair  
$(A_1, B_1)$ and  $(A_2, B_2)$ separately in parallel (using the method described above) and record the results  of the two comparisons in ancilla registers $(\texttt{A}_1\rq{},\texttt{B}_1\rq{})$, $(\texttt{A}_2\rq{}, \texttt{B}_2\rq{})$. The least significant bits of these four ancilla registers can then be used to deduce whether $A>B$ or $A<B$ or $A=B$ with just a single bitwise comparison. 
Thus, we effectively halved the depth by dividing the problem into smaller problems and merging them afterwards.  We now explain a bottom-up implementation.

\begin{figure}[tb]
  \centering
\resizebox{.8\linewidth}{!}{
    \includegraphics{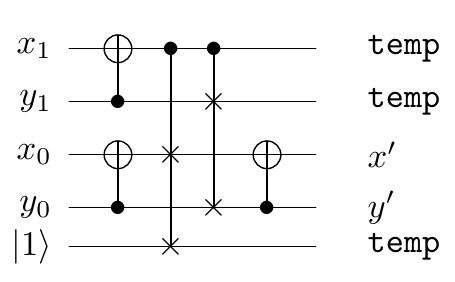}
  }
  \label{fig:Compare2}
  \caption{
    A circuit that implements $\textsc{Compare2}$, taking a pair of 2-bit integers and outputting a pair of single bits while preserving inequalities.
    The input pair is $(x, y) = (x_0 + 2 x_1, y_0 + 2 y_1)$.
    The output pair is $(x^\prime, y^\prime)$ and will satisfy $\text{sign}(x^\prime - y^\prime) = \text{sign}(x - y)$.
    Output qubits marked ``temp" store values that are not needed, and are kept until a later uncompute step where the inputs are restored.
    Each Fredkin gate within the circuit can be computed using 4 T gates and (by storing an ancilla not shown) later uncomputed using 0 T gates \cite{Jones2013,Gidney2017}.
  }
\end{figure}

Instead of comparing the whole registers $\texttt{A}$ and $\texttt{B}$, our parallelized algorithm slices $\texttt{A}$ and $\texttt{B}$ into pairs of bits -- the first slice contains $\texttt{A}[0]$ and $\texttt{A}[1]$, the second slice consists of $\texttt{A}[2]$ and $\texttt{A}[3]$, etc., and in the very same way for $\texttt{B}$. 
The key step takes the corresponding slices of $\texttt{A}$ and $\texttt{B}$ and overwrites the second bit of each slice with the outcome of the comparison.
The first bit of each slice is then ignored, so that the comparison results stored in the second bits become the next layer on which bitwise comparisons are performed. We denote the $i$'th bit forming the registers of the $j^{\text{th}}$ layer by $\texttt{A}^j[i]$ and $\texttt{B}^j[i]$.
The original registers $\texttt{A}$ and $\texttt{B}$ correspond to $j=0$:
$\texttt{A}^0\equiv\texttt{A}$ and $\texttt{B}^0\equiv\texttt{B}$. 
The part of the circuit that implements 
a single bitwise comparison is depicted in \fig{Compare2}.
We denote the corresponding transformation by \lq{}\textsc{Compare2}\rq{}, 
i.e.\ 
$(A^{j+1}[i], B^{j+1}[i]) = 
\textsc{Compare2}(A^j[2i], B^j[2i],  A^j[2i+1], B^j[2i+1])
$, 
meaning that it prepares the bits $A^{j+1}[i], B^{j+1}[i]$ storing the comparison result.

\begin{figure}[tb]
  \centering
  \includegraphics[width=0.5\linewidth]{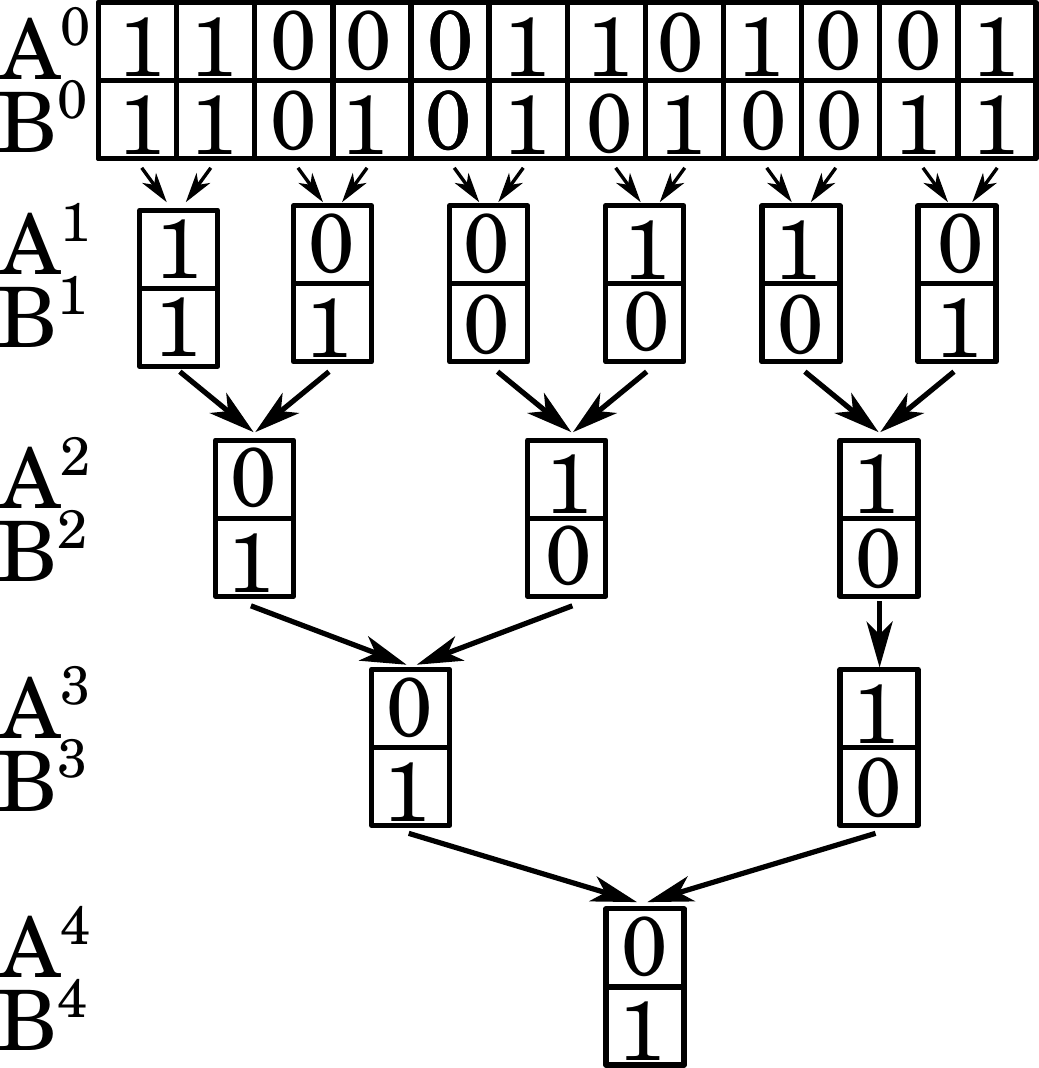}
\caption{Parallelized bitwise comparison. Observe how  
each step reduces the size of the problem by approximately 
one half, while using a constant depth 
for computing the results.}
  \label{fig:ParallelizedComparison}
\end{figure}

At each step, comparisons of the pairs of the original arrays can be performed in parallel, and produce two new arrays with approximately half the size of the original ones to record the results. 
Thus, at each step we approximately halve the size of the problem, 
while using a constant depth 
for computing the results. The basic idea is illustrated in \fig{ParallelizedComparison}.
This procedure is repeated for $\lceil \log d \rceil$ steps\footnote{All logarithms are taken to the base $2$.} 
until registers $\texttt{A}^{\text{\tiny fin}}:=\texttt{A}^{\lceil\log d\rceil}$ and $\texttt{B}^{\text{\tiny fin}}:=\texttt{B}^{\lceil\log d\rceil}$ 
both of size $1$ have been prepared. 

This  parallelized algorithm is perfectly suited 
for comparing arrays whose length $d$ is a power of $2$. 
If $d$ is not a power of $2$,  we can either pad $\texttt{A}$ and $\texttt{B}$ with $0$s prior to their most significant bits 
without altering the result, or introduce comparison of single bits 
(using only the first two gates from the circuit in \fig{Compare2} with targets on $\texttt{A}^{j+1}$ and  $\texttt{B}^{j+1}$ registers respectively).

Formally, we can express our comparison algorithm as follows, 
here assuming $d$ to be a power of $2$:
\begin{algorithmic}
	\For{$j = 0, \ldots, \log d - 1$}
    	\For{$i = 0, \ldots, \text{size}(A^j)/2-1$}
    		\State $\left(A^{j+1}[i], B^{j+1}[i]\right)=\textsc{Compare2}( A^j[2i], B^j[2i],$
            \State $\hspace{3.5cm}  A^j[2i+1], B^j[2i+1])$
      \EndFor
    \EndFor
    \\
    \Return $(A^{\log d - 1}[0], B^{\log d - 1}[0])$
\end{algorithmic}

The key feature of this algorithm is that all the operations of the inner loop can be performed in parallel. Since one application of \textsc{Compare2} requires only constant depth and constant number of operations, our comparison algorithm requires only depth $O(\log d)$. 

\begin{figure}[tb]
  \centering
  \includegraphics[width=.9\linewidth]{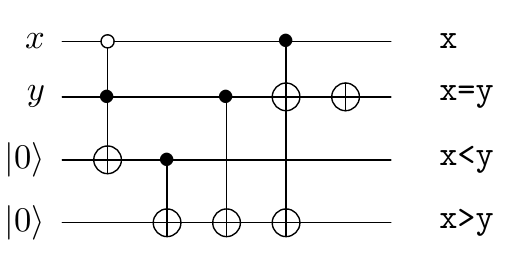}
  \label{fig:Compare2-finish}
  \caption{
    A circuit that determines if two bits are equal, ascending, or descending.
    When the comparison is no longer needed, the results are uncomputed by applying the circuit in reverse order.
  }
\end{figure}

Our comparison algorithm constructed above 
can indeed be used to output a result that 
distinguishes between  $A>B$, $A<B$ and $A=B$. 
Observe that its reversible execution results in 
the ancillary single-qubit registers $\texttt{A}^{\text{\tiny fin}}$ 
and $\texttt{B}^{\text{\tiny fin}}$ generated in the very last step of the algorithm holding information about which number is larger 
or whether they are equal.
Indeed, $\texttt{A}^{\text{\tiny fin}}[0]=\texttt{B}^{\text{\tiny fin}}[0]$
implies $A=B$, $\texttt{A}^{\text{\tiny fin}}[0]<\texttt{B}^{\text{\tiny fin}}[0]$
implies $A<B$, and $\texttt{A}^{\text{\tiny fin}}[0]>\texttt{B}^{\text{\tiny fin}}[0]$ 
implies $A>B$.
The three cases are separated into three control qubits by using the circuit shown in \fig{Compare2-finish}.
These individual control qubits can be used to control further conditional operations that depend on the result of the comparison.

For the purpose in our applications (comparator modules of quantum sorting networks or quantum Fisher-Yates shuffle), we only need to condition on whether $A>B$ is true or false.
Thus, we only need the first operation from the circuit in \fig{Compare2-finish} which takes a single qubit initialized to $\ket{0}$ and transforms it into the output of the comparison oracle.
After the output bit has been produced, we must reverse the complete comparison algorithm (invert the corresponding unitary process), thereby uncomputing all the ancillary registers that have been generated along this reversible process and restoring the input registers $\texttt{A}$ and $\texttt{B}$.

The actual \lq{}comparison oracle\rq{} thus takes as inputs 
two size-$d$ registers $\texttt{A}$ and $\texttt{B}$ (holding values 
$A$ and $B$) and a single-qubit ancilla \texttt{q} initialized 
to $\ket{0}$. It reversibly 
computes whether $A>B$ is true or false by executing the parallelized comparison process presented above. 
It copies the result (which is stored in $\texttt{A}^{\text{\tiny fin}}$)   
to ancilla \texttt{q}. It then executes the inverse of the 
comparison process. 
It outputs $\texttt{A}$ and $\texttt{B}$ {\em unaltered} and the ancilla \texttt{q} 
holding the result of the oracle: $q=1$ if $A>B$ and $q=0$ if $A\le B$. 
As shown, this oracle has circuit size $O(d)$ but depth only $O(\log d)$ and a T-count of $8d + O(1)$.

\section{Symmetrization Using The Quantum Fisher-Yates Shuffle}
\label{app:FisherYatesShuffle}

In this appendix we present an alternative approach for antisymmetrization 
that is not based on sorting, yielding a size- and depth-complexity $O(\eta^2 \log N)$, but with a lower spatial overhead than the sort-based method. Our
alternative symmetrization method uses a quantum variant of the well-known Fisher-Yates shuffle, which applies a permutation chosen uniformly at random to an input array $\texttt{input}$ of length $\eta$. A standard form of the algorithm is given in~\cite{Durstenfeld1964}.

\begin{figure}[bt]
  \centering
  \includegraphics[width=0.87\linewidth]{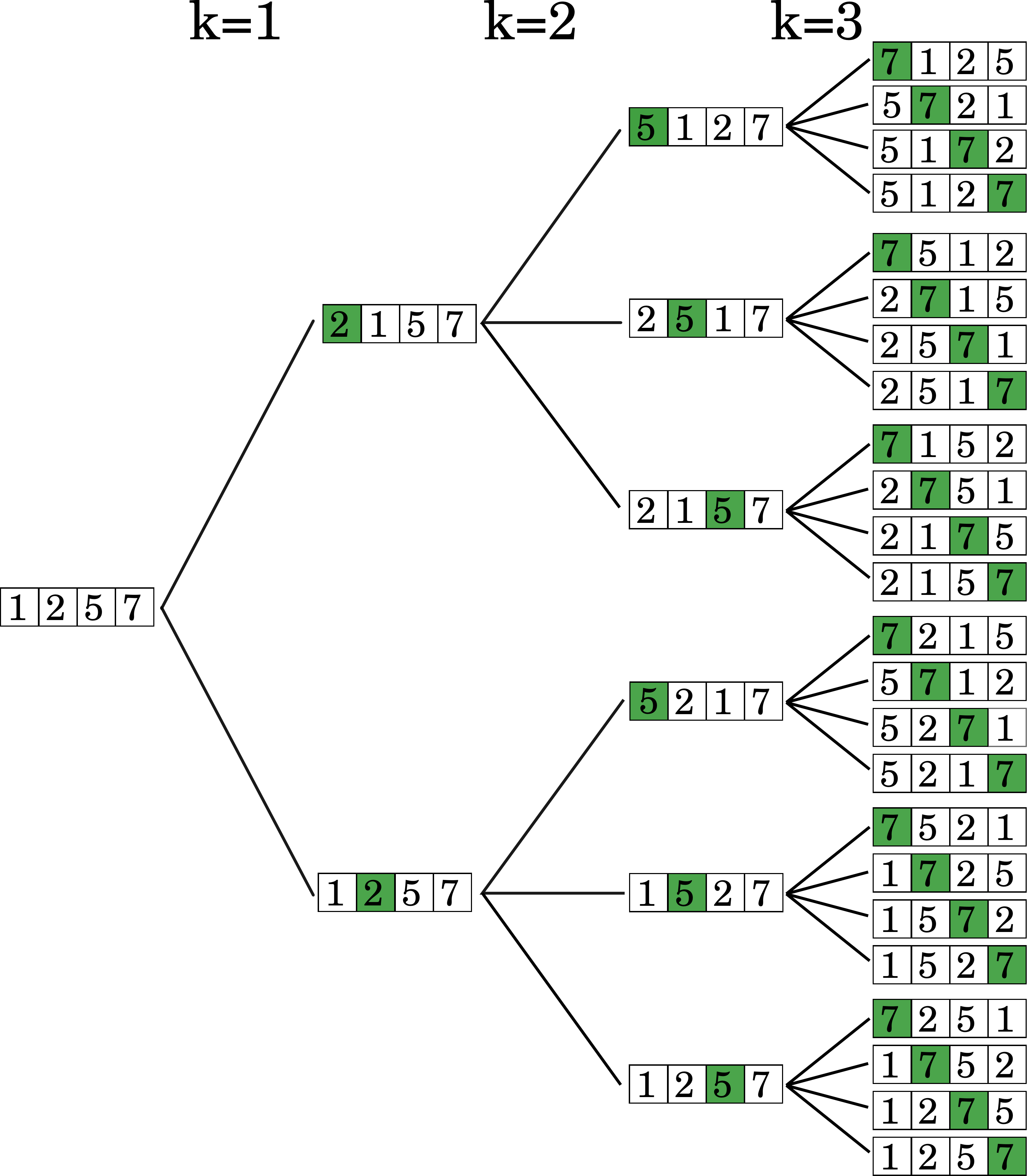}
\caption{A tree diagram for the Fisher-Yates shuffle applied to an example sorted array.  Here the green boxes identify the array entry that has been swapped at each stage of the shuffle.  Observe that the green boxes also label the largest value in the array truncated to position $k$.}
  \label{fig:shuffle}
\end{figure}

We consider the following variant of the Fisher-Yates shuffle: 
\begin{algorithmic}
	\For{$k = 1, \ldots, (\eta - 1)$}
    	\State Choose $\ell$ uniformly at random from $\{0, \ldots, k\}$.
        \State Swap positions $k$ and $\ell$ of \texttt{input}.
    \EndFor
\end{algorithmic}
The basic  idea is illustrated in \fig{shuffle} for $\eta =4$.

There are two key steps that turn the Fisher-Yates shuffle into a quantum algorithm. 
First, our quantum implementation of the shuffle replaces the random selection of swaps with a superposition of all possible swaps.
To achieve this superposition, the random variable is replaced by an equal-weight superposition $\frac{1}{\sqrt{k+1}}\sum_{\ell=0}^k\ket{\ell}$ in an ancillary register (called \texttt{choice}).
At each step of the quantum Fisher-Yates shuffle,
the \texttt{choice} register must begin and end in a fiducial initial state.

In order to reset the \texttt{choice} register, we introduce an additional \texttt{index} register, which initially contains the integers $0,\ldots,\eta-1$.
We shuffle both the length-$\eta$ \texttt{input} register and the \texttt{index} register, and
the simple form of \texttt{index} enables us to easily reset \texttt{choice}.
The resulting state of the joint $\texttt{input}\otimes\texttt{index}$ register is still highly entangled; 
however, provided $\texttt{input}$ was initially sorted in ascending order, we can disentangle $\texttt{index}$ from $\texttt{input}$.

Our quantum Fisher-Yates shuffle consists of the following steps:

\begin{enumerate}
  \item \textbf{Initialization.} \label{step:init}
  Prepare the \texttt{choice} register in the state $\ket{0}$.
  Prepare the \texttt{index} register in the state $\ket{0, 1, \ldots, \eta-1}$.
  Also set a classical variable $k = 1$.
  \item \textbf{Prepare \texttt{choice}.}
  \label{step:prep-choice}
  Transform the \texttt{choice} register from $\ket{0}$ to $\frac{1}{\sqrt{k+1}} \sum_{\ell=0}^{k} \ket{\ell}$.
  \item \textbf{Execute swap.} \label{step:selected-SWAP}
  Swap element $k$ of \texttt{input} with the element specified by \texttt{choice}.
  If a non-trivial swap was executed (i.e.~if \texttt{choice} did not specify $k$),
  apply a phase of $-1$ to the \texttt{input} register.
  Also swap element $k$ of \texttt{index} with the element specified by \texttt{choice}.
  \item \textbf{Reset \texttt{choice}.} \label{step:reset-choice}
  For each $\ell=1, \ldots, k$, subtract $\ell$ from the \texttt{choice} register if position $\ell$ in \texttt{index} is equal to $k$.
  The resulting state of \texttt{choice} is $\ket{0}$.
  \item \textbf{Repeat.}
  Increment $k$ by one.
  If $k < \eta$, go to \step{prep-choice}.
  Otherwise, proceed to the next step.
  \item \textbf{Disentangle \texttt{index} from \texttt{input}.} \label{step:disentangle}
  For each $k \neq \ell = 0, 1, \ldots, \eta-1$, subtract $1$ from position $\ell$ of \texttt{index} if the element at position $k$ in \texttt{input} is greater than the element at position $\ell$ in \texttt{input}.
  The resulting state of \texttt{index} is $\ket{0, 0, \ldots, 0}$, which is disentangled from \texttt{input}.
\end{enumerate}

\begin{figure}[h]
    \centering 
   
	\subfloat[%
        \label{fig:QFY-top}]{  \includegraphics[width=\linewidth]{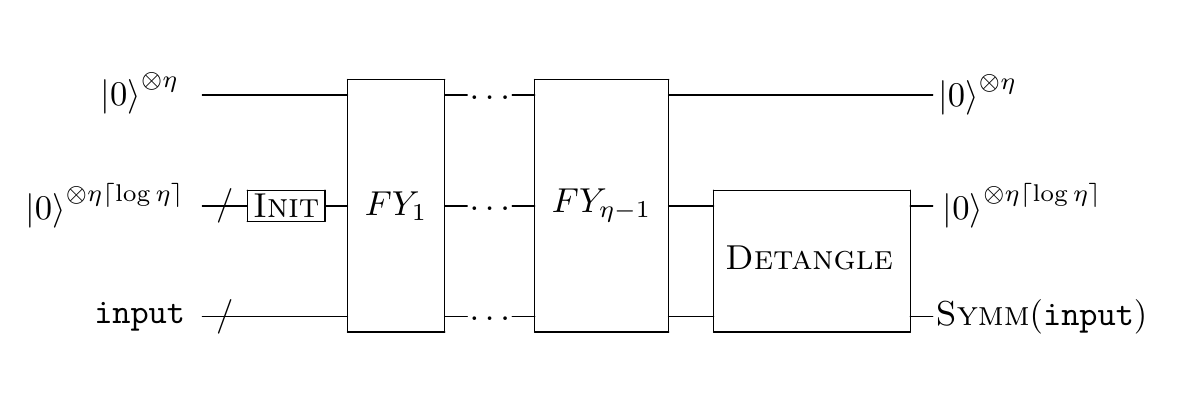}
}
	
	\subfloat[%
    	 \label{fig:FY-block}]{\includegraphics[width=\linewidth]{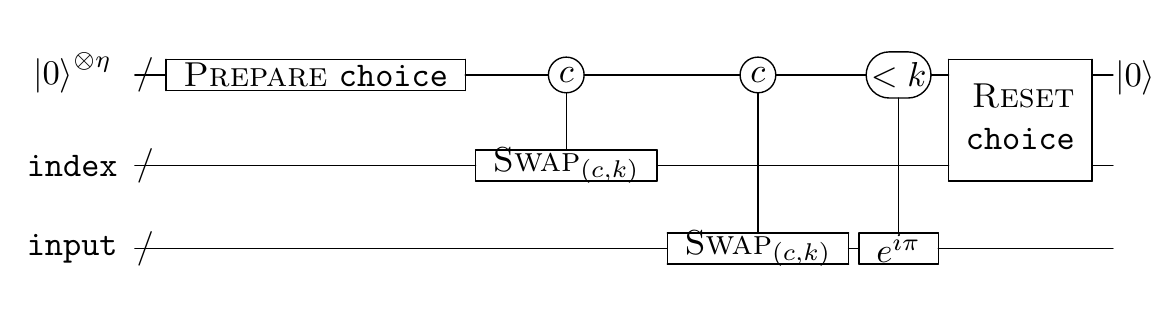}
}
         
	\caption{%
    	An overview of symmetrization by
        quantum Fisher-Yates shuffle.
        (a) High-level view of the algorithm.
        The procedure acts on registers labeled (top to bottom)
        \texttt{choice}, \texttt{index} and \texttt{input}.
        (b) Detail for the Fisher-Yates block $FY_k$.
        The first register (again labeled \texttt{choice}) is used
        to select the target of the two selected swap steps.
        Then a phase $e^{i\pi}=-1$ is applied to the \texttt{input} register if
        a swap was performed, 
        i.e.~if the \texttt{choice} register encodes a value
        less than $k$. Each block $FY_k$ is completed by resetting 
        the \texttt{choice} register back to its original state 
        $\ket{0}^{\otimes \eta}$.
        \label{fig:QFY-overview}}
\end{figure}

We present an overview of the algorithm in \fig{QFY-overview}.
At the highest level, depicted in \fig{QFY-top},
we apply an initialization procedure to \texttt{index},
then $\eta-1$ \lq{}Fisher-Yates\rq{} blocks ($FY_k$ for $k=1,\ldots,\eta-1$),
and finally a disentangling (`\textsc{Detangle}') procedure on \texttt{index} and \texttt{input}.
Following the \textsc{Detangle} procedure, the ancillary registers \texttt{choice} and \texttt{index} are reset to their initial all-zero states
and the \texttt{input} register has been symmetrized.
In each Fisher-Yates block, depicted in \fig{FY-block},
we apply the preparation operator $\Pi_k$ to \texttt{choice},
apply selected swaps on \texttt{choice}+\texttt{index} and \texttt{choice}+\texttt{input},
then apply a phase conditioned on \texttt{choice} to \texttt{input},
and finally reset the \texttt{choice} register.
Preparing and resetting $\texttt{choice}$ as well as executing swap are therefore part of each Fisher-Yates 
block and are thus each applied a total of $\eta - 1$ times (for each of $k = 1,\dots,\eta-1$). Their gate counts and circuits 
depths must thus be multiplied by $(\eta - 1)$. 
Disentangling \texttt{index} and \texttt{input} 
is the most expensive step, but it is executed only once, 
so it contributes only an additive
cost to the overall resource requirement.

In what follows,  we explain each step of the algorithm and justify their 
corresponding resource contributions, which are briefly summarized here: 
\step{init} requires $O(\eta\log\eta)$ gates 
but has a negligible depth $O(1)$.
\step{prep-choice} requires $O(\eta)$ gates 
and has the same depth complexity.
\step{selected-SWAP} requires $O(\eta\log N)$ gates and has also depth $O(\eta\log N)$.
\step{reset-choice} requires $O(\eta \log \eta)$ gates but has only depth $O(\log \eta)$.
As \step{prep-choice} to \step{reset-choice} are repeated $\eta-1$ times, 
the total gate count before \step{disentangle} is $O(\eta^2 \log N)$.
Finally, \step{disentangle} requires $O(\eta^2 \log N)$ gates and has 
depth $O\left(\eta^2\left[\log\log N+\log\eta\right]\right)$.
Thus the total gate count of the quantum Fisher-Yates shuffle is $O(\eta^2 \log N)$. Because most of the gates need to be performed sequentially, the overall circuit depth of the algorithm is also $O(\eta^2 \log N)$.

Our complexity analysis is given in terms of elementary gate operations,
a term we use loosely.
Generally speaking, we treat all single-qubit gates as elementary and we allow up to two controls for free on each single-qubit gate.
This definition of elementary gates includes several standard universal gate sets such as Clifford+T and Hadamard+Toffoli.
A more restrictive choice of elementary gate set only introduces somewhat larger constant factors in most of the procedure.
The exception is the application of $\Pi_k$ in the first step of $FY_k$,
where we require the ability to perform controlled single-qubit rotations of angle
$\arcsin \left( \sqrt{\frac{\ell}{\ell+1}} \right)$,
where $\ell = 1, \ldots, k$.
The Solovay-Kitaev theorem implies a gate-count overhead that grows polylogarithmically in the inverse of the error tolerance. We now proceed by analyzing each step to the quantum Fisher-Yates shuffle.

\subsection{Initialization}
\label{sec:QFY-init}

The first step  is to initialize \texttt{choice} in the state $\ket{0}^{\otimes \eta}$. This is assumed to have zero cost. 
The \texttt{index} register is set to the state $\ket{0, 1, \ldots, \eta-1}$ that represents the positions of the entries of \texttt{input} in ascending order. Because each of the $\eta$ entries in \texttt{index} must be capable of storing any of the values $0, 1, \ldots, \eta-1$, the size of \texttt{index} is $\eta \lceil \log \eta \rceil$ qubits. This step requires $O (\eta\log\eta)$ single-qubit gates that can be applied in parallel with circuit depth $O(1)$.

\subsection{Fisher-Yates Blocks}
\label{sec:FY-block}

Each Fisher-Yates block has three stages:
prepare \texttt{choice},
executed selected swaps, and
reset \texttt{choice}.
The exact steps depend on the encoding of the \texttt{choice} register;
in particular, whether it is binary or unary.

We elect the conceptually simplest encoding of \texttt{choice},
which is a kind of unary encoding.
We use $\eta$ qubits (labelled $0,1,\ldots,\eta$), define
\begin{equation}
\ket{\texttt{null}} = \ket{0}^{\otimes\eta}
\end{equation}
and encode
\begin{equation}
\ket{\ell} = X_\ell \ket{\texttt{null}},
\end{equation}
where $X_\ell$ is the Pauli $X$ applied to the qubit labelled $\ell$.

An advantage of our encoding for \texttt{choice} is that the selected swaps require only single-qubit controls.
An obvious disadvantage is the unnecessary space overhead.
Although one can save space with a binary encoding,
the resulting operations become somewhat more complicated and hence come at an increased time cost.
Our choice of encoding is made for simplicity.

\subsubsection{Prepare \texttt{choice}}
\label{sec:PrepareChoice}

\begin{figure}[bt]
\includegraphics[width=\linewidth]{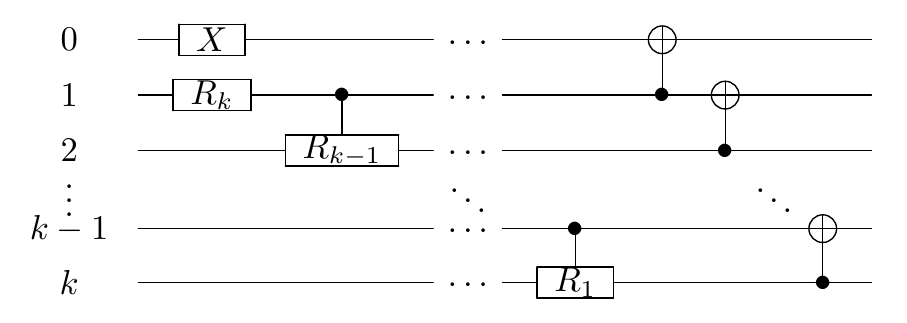}
 \caption{%
 Circuit for preparing the \texttt{choice} register at the beginning of block $FY_k$.
 See \eq{rot_mtx} for the definition of $R_\ell$.} 
 \label{fig:Prepare-choice}
  \end{figure}

Our preparation procedure has two stages.
First, we prepare an alternative unary encoding of the state
\begin{equation}
\ket{W_k} :=
\frac{1}{\sqrt{k+1}} \sum_{\ell=0}^k \ket{\ell},
\end{equation}
which we name for its resemblance to the W-state $\frac{1}{\sqrt{3}} (\ket{001} + \ket{010} + \ket{100})$.
Second, we translate the alternative unary encoding to our desired encoding.
For a summary of the procedure, see \fig{Prepare-choice}. 

Next, we explain how to prepare $\ket{W_k}$ in the alternative unary encoding.
The alternative encoding is
\begin{equation}
\ket{\ell} =
\left(\prod_{\ell^\prime=0}^\ell X_{\ell^\prime} \right) \ket{\texttt{null}}.
\end{equation}
We can prepare $\ket{W_k}$ in this encoding with a cascade of controlled rotations of the form 
\begin{equation}
\label{eq:rot_mtx}
R_\ell :=
\frac{1}{\sqrt{\ell+1}}
\begin{pmatrix}
	1 & -\sqrt{\ell}\\
	\sqrt{\ell} & 1
\end{pmatrix}.
\end{equation}
Explicitly:
\begin{algorithmic}
	\State Apply $X$ to qubit 0.
	\State Apply $R_k$ to qubit 1.
	\For{$\ell = 1, \ldots, k-1$}
		\State Apply $R_{k-\ell}$ controlled on qubit $\ell$ to qubit $\ell+1$.
	\EndFor
\end{algorithmic}
This is a total of $k+1$ gates,
$k=1$ of which are applied sequentially.

Next we explain how to translate to the desired encoding.
This is a simple procedure:
\begin{algorithmic}
	\For{$\ell = k, \ldots, 1$}
		\State Apply \textsc{Not} controlled on qubit $\ell$ to qubit $\ell-1$.
	\EndFor
\end{algorithmic}
The total number of \textsc{CNot} gates is $k$,
and they must be applied in sequence.
Thus the total gate count (and time-complexity) for preparing \texttt{choice} is $O(k) = O(\eta)$.

\subsubsection{Selected Swap}
\label{sec:SelSwap}

\begin{figure}[h]
 \centering
\includegraphics[width=\linewidth]{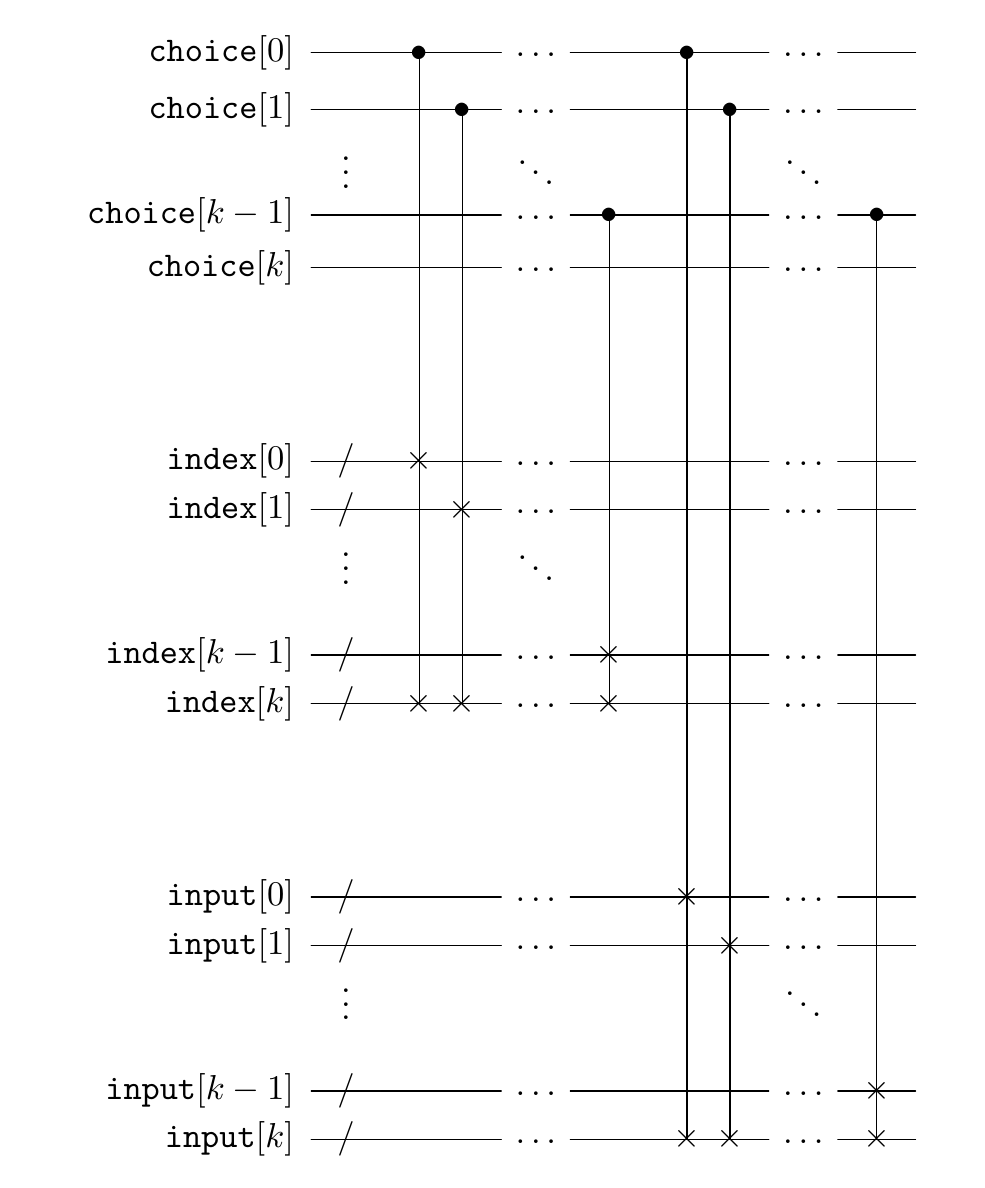}
\caption{Implementation of the two selected swaps $\textsc{SelSwap}_k$ as part of $FY_k$,  with the unary-encoded \texttt{choice} as the control register and \texttt{index} and \texttt{input} as target registers, respectively.
As each wire of the target registers stand for several qubits, each controlled-\textsc{Swap} is to be interepreted as many bitwise controlled-\textsc{Swap}s.}
\label{fig:Circuit-SelSwap}
  \end{figure}

We need to implement selected swaps of the form 
\begin{equation}
\textsc{SelSwap}_k :=
	\sum_{c=0}^{\eta-1}
	\proj{c}_\texttt{choice} \otimes
    \textsc{Swap}(c,k)_\texttt{target}, 
\end{equation}
where the $\textsc{Swap}(c,k)$ operator acts on either
$\texttt{target} = \texttt{index}$ or $\texttt{target} = \texttt{input}$.
Here the state of the \texttt{choice} register {\em selects} which entry in the \texttt{target} array
is to be swapped with entry $k$.
Our unary encoding of the \texttt{choice} register allows for a simple implementation of \textsc{SelSwap};
see \fig{Circuit-SelSwap}.

Observe that only the first $k+1$ subregisters are involved of each \texttt{choice}, \texttt{index} and \texttt{input}, respectively.
Also observe that, for each $i=0,1,\dots,k$,
$\texttt{index}[i]$ is of size $\lceil\log\eta\rceil$ whereas $\texttt{input}[i]$ is of size $\lceil\log N\rceil$.
Hence, the circuit actually consists of
$k \lceil\log \eta\rceil + 
k \lceil\log N \rceil$ ordinary 3-qubit controlled-\textsc{Swap} gates that for the most part must be executed sequentially. 
As $\eta \leq N$, we report $O \left(\eta \log N\right)$ 
for both gate count and depth. 

\subsubsection{Applying the controlled-phase}

Applying the controlled-phase gate is straightforward.
We select a target qubit in the \texttt{input} register -- it does not matter which.
Then, for each $\ell = 0, 1, \ldots, k-1$, we apply a phase gate controlled on position $\ell$ of \texttt{choice} to the target qubit.
The result is that \texttt{input} has picked up a phase of $(-1)$ if \texttt{choice} specified a value strictly less than $k$.
The total number of gates is $k = O(\eta)$, while the depth can be made $O(1)$. 

\subsubsection{Resetting \texttt{choice} register}
\label{sec:ResetChoice}

\begin{figure}[h]
  \centering
\hspace{-2cm}\includegraphics[width=.8\linewidth]{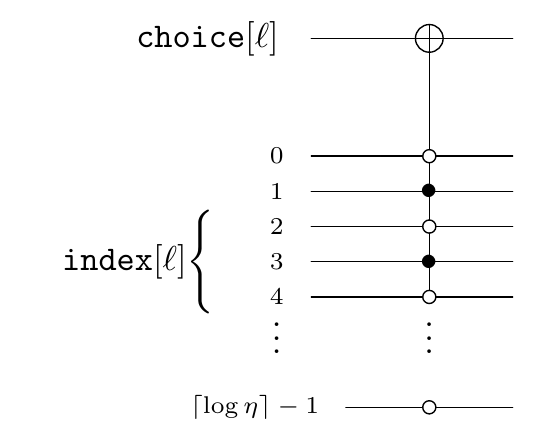}
\caption{
Circuit for resetting $\texttt{choice}$ 
register as part of iteration block $FY_k$. In this example $k=10$. It consists of a series of multi-fold-controlled-\textsc{Not}s, employing the $\ell$-th wire of $\texttt{choice}$  and the $\ell$-th subregister $\texttt{index}[\ell]$ of size $\lceil\log \eta\rceil$,  
for each $\ell = 0, \dots, k$.
Note that the multi-fold-controlled-\textsc{Not} is the same for all values of $\ell$. The control sequence is a binary encoding of $k = 10$. The \textsc{Not} erases $\texttt{choice}[\ell]$ if $\texttt{index}[\ell] = k$.} \label{fig:Reset-Choice}
\end{figure}

The reason we execute swaps on both \texttt{index} and \texttt{input} is to enable reversible erasure of \texttt{choice} at the end of each Fisher-Yates block.
This is done by scanning \texttt{index} to find out which value of $k$ was encoded into \texttt{choice}.
In general, we know that step $k$ sends the value $k$ to position $\ell$ of \texttt{index},
where $\ell$ is specified by the \texttt{choice} register.
We thus erase \texttt{choice} by applying a \textsc{Not} operation to $\texttt{choice}[\ell]$ if $\texttt{index}[\ell] = k$.
This can be expressed as a multi-controlled-\textsc{Not}, as illustrated by an example in \fig{Reset-Choice}.
The control sequence of the multi-controlled-\textsc{Not} is a binary encoding of the value $k$. 

For compiling multiple controls, see Figure 4.10 in \cite{NielsenChuang}. 
Each $\lceil\log \eta\rceil$-fold-controlled-\textsc{Not} 
can be decomposed into a network of $O(\log \eta)$ gates (predominantly Toffolis) with depth $O(\log \eta)$. Because the $k+1$ multi-fold-controlled-\textsc{Not}s (for $\ell\le k\le \eta -1$) can all be executed in parallel, resetting $\texttt{choice}$ register thus requires a circuit 
with $O(\eta \log \eta)$ gates but only $O(\log \eta)$ depth.

\subsection{Disentangling \texttt{index} from \texttt{input}}
\label{sec:FY-detangle}
The last task is to clean up and disentangle \texttt{index} from \texttt{input} by resetting the former to the 
original state $\ket{0}^{\otimes \eta\lceil\log \eta\rceil}$ while leaving the latter in the desired antisymmetrized superposition. This can be achieved as follows. 

We compare the value carried by each of the $\eta$ subregisters \texttt{input$[\ell]$} 
(labeled by position index $\ell=0,1,\dots,\eta-1$) 
with the value of each other subregister \texttt{input$[\ell']$} ($\ell'\not=\ell$), 
thus requiring $\eta(\eta-1)$ comparisons in total. 
Note that these subregisters of \texttt{input} have all size $\lceil\log N\rceil$. 
Each time the value held in \texttt{input$[\ell]$} 
is larger than the value carried by any other of the remaining $\eta-1$ subregisters 
\texttt{input$[\ell']$}, we {\em decrement} the value of the corresponding $\ell^{\mbox{\tiny th}}$ subregister \texttt{index$[\ell]$} 
of \texttt{index} {\em by $1$}. In cases in which the value carried by 
\texttt{input$[\ell]$} is smaller than \texttt{input$[\ell']$}, we do not decrement 
the value of \texttt{index$[\ell]$}.  After accomplishing all the $\eta(\eta-1)$  
comparisons within the \texttt{input} register and controlled
decrements, we have reset
the \texttt{index} register state to 
$\ket{0}^{\otimes \eta\lceil\log \eta\rceil}$ 
while leaving the \texttt{input} 
register in the antisymmetrized superposition state. 

Each comparison between the values of two subregisters of  \texttt{input} (each of size $\lceil\log N\rceil$) can be performed using the comparison oracle introduced in~\app{ComparisonOracle}. 
The oracle\rq{}s output is then used to control the 
\lq{}{\em decrement by 1}\rq{} operation, 
after which the oracle is used again 
to uncompute the ancilla holding its result.  
The comparison oracle has been shown to require $O(\log N)$ gates 
but to have only circuit depth $O(\log \log N)$. 

\begin{figure}[tbh]
  \centering
\begin{align}
\includegraphics[width=.9\linewidth]{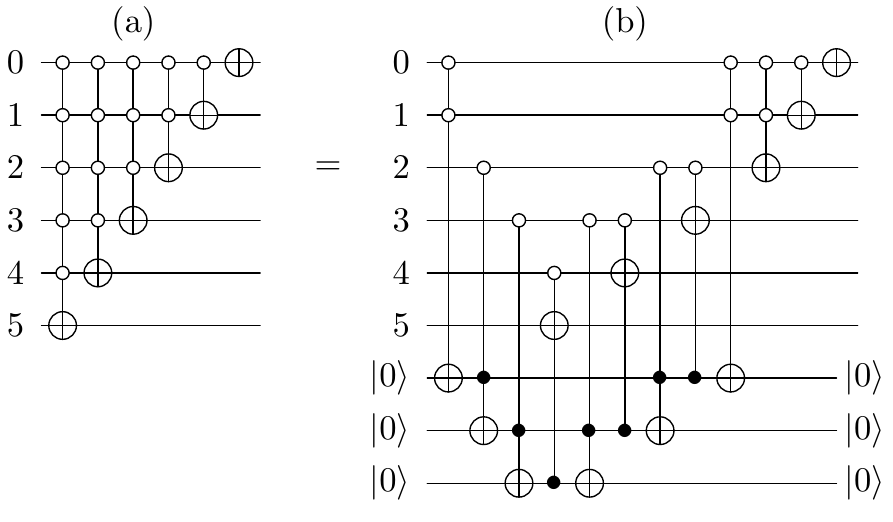}\nonumber
\end{align}
  \label{fig:Decrement-by-1}
  \caption{Circuit implementing \lq{}{\em decrement by 1}\rq{} operation, applied to \texttt{index}$[\ell]$ subregisters of size $\lceil\log\eta\rceil$.  
 (a) Example for $\eta =64$. (b) Decomposition into a network of $O\left(\log\eta\right)$ Toffoli gates  
  using $O\left(\log\eta\right)$ ancillae.}
\end{figure}

Decrementing the value of the $\lceil\log\eta\rceil$-sized index subregister \texttt{index$[\ell]$} (for any $\ell=0,1,\dots, \eta -1$) by the value 1 
can be achieved by a circuit depicted in \fig{Decrement-by-1}.

Each such operation involves a total of $\lceil\log\eta\rceil$ multi-fold-controlled-\textsc{Not}s. More specifically, it involves $n$-fold-controlled-\textsc{Not}s for each 
$n=\lceil\log\eta\rceil-1,\dots,0$. Note 
that each must also be controlled by the qubit holding the result of the 
comparison oracle. 
When decomposing each of them into a network 
of $O(n)$ Toffoli gates using $O(n)$ ancillae according to the method provided in Figure 4.10 in \cite{NielsenChuang}, 
the majority of the involved Toffoli gates for different 
values of $n$ effectively cancel each other out. The resulting cost is 
only $O\left(\log\eta\right)$ Toffolis rather than $O\left(\log^2\eta\right)$, at the expense of an additional space overhead of size $O\left(\log \eta\right)$. However, there is no need to employ new ancillae.
We can simply reuse those qubits that previously composed the \texttt{choice} register for this purpose, 
as the latter is not being used otherwise 
at this stage any more.

Putting everything together, 
the overall circuit size for this step  
amounts to $O\left(\eta(\eta-1)\left[\log N+\log\eta\right] \right)$ predominantly Toffoli gates, 
which can then be further decomposed into 
\textsc{CNot}s and single-qubit gates (including $T$ gates) 
in well-known ways.
Because $\eta\le N$, 
we thus report $O(\eta^2 \log N)$ for the overall 
gate count for this step, while its circuit depth 
is $O\left(\eta^2\left[\log\log N+\log\eta\right]\right)$.

\end{document}